\documentclass{aa}
\usepackage[varg]{txfonts}
\usepackage{natbib}
\usepackage{graphicx}

\usepackage{color}

\bibpunct{(}{)}{;}{a}{}{,}

\begin{document}

\title{Mapping accretion and its variability in the young open cluster NGC 2264: a study based on u-band photometry\thanks{Based on observations obtained with MegaPrime/MegaCam, a joint project of CFHT and CEA/DAPNIA, at the Canada-France-Hawaii Telescope (CFHT) which is operated by the National Research Council (NRC) of Canada, the Institut National des Sciences de l'Univers of the Centre National de la Recherche Scientifique (CNRS) of France, and the University of Hawaii.}\fnmsep\thanks{Tables 2, 3 and 4 are only available in electronic form at the CDS via anonymous ftp to cdsarc.u-strasbg.fr (130.79.128.5) or via http://cdsweb.u-strasbg.fr/cgi-bin/qcat?J/A+A/}}

\author{L. Venuti\inst{1,2,3} \and J. Bouvier\inst{1,2} \and E. Flaccomio\inst{4} \and S.\,H.\,P. Alencar\inst{5} \and J. Irwin\inst{6} \and J.\,R. Stauffer\inst{7} \and A. M. Cody\inst{7} \and P.\,S. Teixeira\inst{8} \and A.\,P.~Sousa\inst{5} \and G.~Micela\inst{4} \and J.-C. Cuillandre\inst{9} \and G. Peres\inst{3}}

\institute{Univ. Grenoble Alpes, IPAG, F-38000 Grenoble, France \and CNRS, IPAG, F-38000 Grenoble, France \\e-mail: Laura.Venuti@obs.ujf-grenoble.fr \and Dipartimento di Fisica e Chimica, Universit\`a degli Studi di Palermo, Piazza del Parlamento 1, 90134 Palermo, Italy \and Istituto Nazionale di Astrofisica, Osservatorio Astronomico di Palermo G.S. Vaiana, Piazza del Parlamento 1, 90134 Palermo, Italy \and Departamento de F\'isica - ICEx - UFMG, Av. Ant\^onio Carlos, 6627, 30270-901 Belo Horizonte, MG, Brazil \and Harvard-Smithsonian Center for Astrophysics, 60 Garden Street, Cambridge, MA 02138, USA \and Spitzer Science Center, California Institute of Technology, 1200 E California Blvd., Pasadena, CA 91125, USA  \and Universit\"at Wien, Institut f\"ur Astrophysik, T\"urkenschanzstrasse 17, 1180, Vienna, Austria \and Canada - France - Hawaii Telescope Corporation, 65-1238 Mamalahoa Highway, Kamuela, HI, 96743, USA}

\date{Received 7 March 2014 / Accepted 29 July 2014}

\abstract {The accretion process has a central role in the formation of stars and planets.}{We aim at characterizing the accretion properties of several hundred members of the star-forming cluster NGC~2264 (3~Myr). }{We performed a deep {\it ugri} mapping as well as a simultaneous $u$-band+$r$-band monitoring of the star-forming region with CFHT/MegaCam in order to directly probe the accretion process onto the star from UV excess measurements. Photometric properties and stellar parameters are determined homogeneously for about 750 monitored young objects, spanning the mass range $\sim$0.1--2\,M$_\odot$.  About 40\% of the sample are classical (accreting) T Tauri stars, based on various diagnostics (H$_\alpha$, UV and IR excesses). The remaining non-accreting members define the (photospheric + chromospheric) reference UV emission level over which flux excess is detected and measured.}{We revise the membership status of cluster members based on UV accretion signatures, and report a new population of 50 CTTS candidates. A large range of UV excess is measured for the CTTS population, varying from a few times 0.1 to $\sim$3 mag. We convert these values to accretion luminosities and accretion rates, via a phenomenological description of the accretion shock emission. We thus obtain mass accretion rates ranging from a few $10^{-10}$ to $\sim$$10^{-7}$~M$_\odot$/yr. Taking into account a mass-dependent detection threshold for weakly accreting objects, we find a $>$6$\sigma$ correlation between mass accretion rate and stellar mass. A power-law fit, properly accounting for censored data (upper limits), yields $\dot{M}_{acc}\propto$ M$_*^{1.4\pm0.3}$. At any given stellar mass, we find a large spread of accretion rates, extending over about 2 orders of magnitude. The monitoring of the UV excess on a timescale of a couple of weeks indicates that its variability typically amounts to 0.5~dex, i.e., much smaller than the observed spread in accretion rates. 
We suggest that a non-negligible age spread across the star-forming region may effectively contribute to the observed spread in accretion rates at a given mass. In addition, different accretion mechanisms (like, e.g.,  short-lived accretion bursts vs. more stable funnel-flow accretion) may be associated to different $\dot{M}_{acc}$ regimes.}{A huge variety of accretion properties is observed for young stellar objects in the NGC~2264 cluster. While a definite correlation seems to hold between mass accretion rate and stellar mass over the mass range probed here, the origin of the large intrinsic spread observed in mass accretion rates at any given mass remains to be explored.}

\keywords{Accretion, accretion disks - Stars: formation - Stars: low-mass - Stars: pre-main sequence - open clusters and associations: individual: NGC 2264 - Ultraviolet: stars}

\titlerunning{Mapping accretion and its variability in the young open cluster NGC 2264}
\maketitle

\section{Introduction}
The disk accretion phase assumes a pivotal role in the scenarios of early stellar evolution and planetary formation. Circumstellar disks are the ubiquitous result of the earliest stages of star formation, surrounding the vast majority of solar-type protostars at an age of $\sim$1 Myr. Mass accretion from the inner disk to the central star regulates the star-disk interaction over the few subsequent million years; this process has a direct, long-lasting impact on the basic properties of the system, by providing at the same time mass and angular momentum. Angular momentum transport and mass infall have a most important part in dictating the dynamical evolution of protoplanetary disks, hence setting the environmental conditions which eventually lead to the formation of planetary systems \citep{armitage2011}. 

The widely accepted paradigm for mass accretion in young stellar objects (YSOs) is that of magnetospheric accretion \citep{camenzind1990, konigl1991}. A cavity of a few stellar radii extends from the stellar surface to the inner disk rim, as an effect of the inner disk truncation by the stellar magnetosphere (B$\sim$2~kG). Accretion columns, channeled along the magnetic field lines, regulate the transport of material from the disk inner edge to the central star; as the accreting material impacts onto the stellar surface at near free-fall velocity, accretion shocks are produced near the magnetic poles. This picture is strongly supported by its predictive capability of the main observational features of accreting young solar-type stars (classical T Tauri stars, or CTTS): UV, optical and IR excess compared to the photospheric flux; spectral ``veiling''; broad emission lines; inverse P Cygni profiles; pronounced spectroscopic and photometric variability \citep[e.g.][]{bouvier2007}. 

Several studies \citep[e.g.][]{muzerolle2003, rigliaco2011} have shown that the rates ($\dot{M}_{acc}$) at which mass accretion occurs tend to scale with the mass of the central object (M$_*$). Notwithstanding this general trend, a large spread in the $\dot{M}_{acc}$ values at a given M$_*$ appears to be a constant feature \citep[e.g.][]{hartmann2006}; this is indicative of a complex and dynamic picture, shaped by a variety of parameters and concurrent processes. Similarly, observations suggest a progressive decrease of $\dot{M}_{acc}$ over the age of the systems \citep{hartmann1998, sicilia_aguilar2010}, following the time evolution and dispersal of circumstellar disks \citep{haisch2001, hillenbrand2005}. The large spread associated with the effective relationship attests that a multiplicity of elements concur to the disk dissipation; moreover, the initial conditions in which stars are formed \citep[e.g.][]{dullemond2006} and local environmental properties \citep[e.g.][]{guarcello2007} are likely to significantly affect the evolutionary pattern of individual star--disk systems.

Statistical surveys of the accretion properties of young stars provide essential insight into the dynamics of accretion. Large, homogeneously characterized samples of objects, spanning several ranges of parameters, are crucial to achieve a proper understanding of the physics governing the disk accretion history of forming stars and to explore different facets of the accretion mechanisms. To this respect, extensive mapping surveys of well-populated star-forming regions at UV wavelengths \citep[e.g.][]{rigliaco2011, manara2012} provide a most interesting standpoint, as they allow to directly probe accretion for large samples of stars from the excess emission produced in the shocked impact layer at the base of the accretion column. 

Such single epoch surveys efficiently trace the instantaneous spread of accretion properties in a given star-forming region; however, they are unable to probe the nature of such a spread, i.e. to determine whether this is due to variability of individual objects or to an intrinsic spread amongst objects. And yet, a key aspect of the accretion dynamics is the intrinsic variability profile of the process (i.e., the amplitude and characteristic timescales of the variability in $\dot{M}_{acc}$). The enhanced variability typical of young stars, cornerstone of the very definition of the T Tauri class \citep{joy1945}, is manifest over a broad range of wavelengths (X-rays, UV, optical, IR) and of time baselines (hours, days, years); a variety of processes concur to shape the variability pattern of individual objects, with a significant contribution coming from the geometry of the system. Understanding to which extent and on which timescales the accretion process is intrinsically variable, and characterizing its variability on a statistical basis compared to individual cases, is of utmost importance in order to accurately interpret the information provided by large accretion surveys and thus infer a detailed picture of the accretion process. 

Multi-epoch surveys hence provide a more accurate description of the accretion process, as they allow to probe the time evolution of single snapshots of the accretion properties and, thus, to investigate how accretion occurs. This has been addressed in a few recent studies \citep[e.g.][]{nguyen2009, fang2013}, which explored the variability of spectral accretion diagnostics for tens of objects with a few observing epochs over several months. Such efforts provided very valuable constraints on the average accretion variability shown by typical CTTS. However, a much tighter time sampling is needed in order to recover an estimate of the effective intrinsic accretion variability, isolated from geometric contributions which are simply due to a varying visible portion of the accretion spots during stellar rotation.

We have recently conducted an extensive observational program specifically aimed at addressing the issue of YSOs variability over the full spectrum. The \textit{Coordinated Synoptic Investigation of NGC~2264} (CSI~2264; \citealp{C13CSI2264}) project was devised as a coordinated multi-wavelength exploration of the hours-to-weeks variability of the pre-Main Sequence (PMS) population of the star-forming region NGC~2264. The space telescopes CoRoT and Spitzer provided the backbone of the optical and IR investigations, with simultaneous continuous monitoring over 40 and 30 days respectively; the Flames multi-object spectrograph at the Very Large Telescope (VLT) provided a full set of spectra covering 20 different epochs for $\sim$100 young stars, while multi-band optical+UV photometric monitoring obtained at the Canada-France-Hawaii Telescope (CFHT) provided contemporaneous, synoptic measures of the accretion rates and their variability from the direct diagnostics of the UV excess, with an overall time coverage of 14 consecutive days and several measurements per night. The targeted region, NGC~2264, has long been a benchmark for star formation studies, thanks to its youth ($\sim$3~Myr), its richness ($\sim$1500 known members, both CTTS and non-accreting, weak-lined T Tauri stars, or WTTS), its relative proximity (d$\sim$800 pc), the low extinction on the line of sight and the association with a molecular cloud complex that significantly reduces the contamination from background stars (see \citealp{dahm08} for a recent review on the region).

As part of the CSI~2264 campaign, this paper specifically focuses on the issue of accretion properties and accretion variability in the PMS population of NGC~2264. Sect.\,\ref{sec:data} describes the photometric dataset obtained at CFHT, on which the study reported here is based. Sect.\,\ref{sec:population} reports on the characterization of photometric accretion signatures, which allowed us to perform a new, accretion-driven census of NGC~2264 members; an extensive, homogeneous investigation of individual stellar properties is also reported. In Sect.\,\ref{sec:accretion}, we describe the procedure adopted to convert the UV excess measurement to a $\dot{M}_{acc}$ estimate; we analyze the $\dot{M}_{acc}$ over M$_*$ distribution thus inferred and compare this picture with the amount of variability registered over a baseline of two weeks; we further probe the origin of this variability and infer a specific estimate of the intrinsic accretion variability as opposed to rotational modulation detected during the monitoring. Results are discussed in Sect.\,\ref{sec:discussion}; our conclusions are synthetized in Sect.\,\ref{sec:conclusions}. This provides a complete picture of a whole star-forming region and its several hundred members at short wavelengths, from a specifically accretion-oriented perspective. A more detailed characterization of the variability properties monitored at CFHT and their relation to the underlying physics will be addressed in a forthcoming paper (Venuti et al., in preparation).

\section{Observations and data reduction} \label{sec:data}

\subsection{CFHT MegaCam surveys of NGC 2264: data reduction and photometric calibration}
We surveyed NGC 2264 in two multiwavelength observing campaigns at the CFHT, using the wide-field optical camera MegaCam; the camera has a field of view (FOV) of $\sim$1~deg.$^2$ \citep{boulade03}, thus fitting the whole region in a single telescope pointing. Five broad-band filters ($u^*, g', r', i', z'$) are adopted at the camera; their design, close to the Sloan Digital Sky Survey (SDSS) photometric system \citep{F96SDSS}, ensures high efficiency for faint object detection and deep sky mapping.

\subsubsection{Mapping survey}\label{sec:mapping}

The first MegaCam survey of NGC 2264, held on Dec. 12, 2010, consisted of a deep $u^*g'r'$ mapping of the region. In the $r'$-band, five short exposures (10 s) were obtained, using a dithering pattern in order to compensate for the presence of bad pixels and small gaps on the CCD mosaic. In the $u^*$-band and $g'$-band, the same 5-step dithering procedure was repeated in two different exposure modes, short (10 s) and long (60 s), in order to obtain a good signal/noise (S/N) for the whole sample of objects without saturating the brightest sources. All images were obtained over a continuous temporal sequence within the night. On Feb. 28, 2012, concurrently with the monitoring survey reported in Sect.\,\ref{sec:monitoring}, we additionally obtained a deep $i'$-band mapping, in order to reconstruct a full 4-band picture of NGC~2264 together with the $u^*g'r'$ survey of Dec. 2010. Similarly to previous observations, a dithering pattern was used, along with two different exposure times: 180 s, to detect all sources in the FOV, and 5 s, to recover the bright sources saturated in the long exposure.

\begin{figure}
\resizebox{\hsize}{!}{\includegraphics{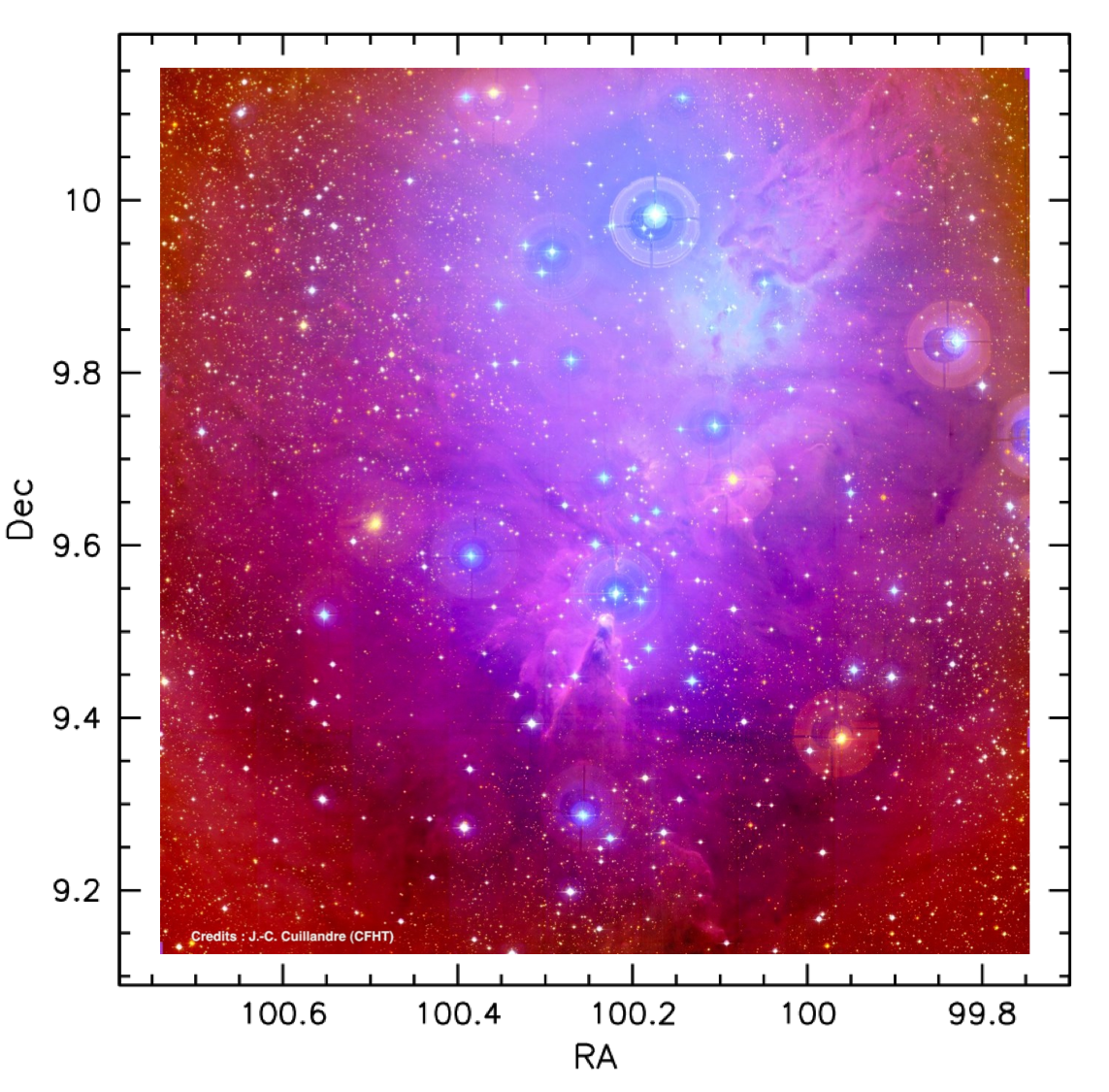}}
\caption{CFHT/MegaCam $ugr$ color image of NGC 2264.}
\label{fig:NGC2264}
\end{figure}

Raw data have been preliminarily processed at the CFHT, in order to correct for instrumental signatures across the whole CCD mosaic (i.e., bad pixel clusters or columns, dark current, bias, non-uniformity in the response), and a first astrometric and photometric calibration has been performed. The subsequent photometric processing of the images obtained during the survey has been handled as described below.

For each spectral band and each exposure time, single dithered images have been combined, in order to discard spurious signal detections (e.g., cosmic rays) and correct for small CCD gaps. The photometric images thus assembled (one in the $r'$-band and two for each other filter, corresponding to the two different exposure times) have then been processed using the \mbox{SExtractor} \citep{sextractor} and \mbox{PSFEx} \citep{psfex} tools, in order to identify true sources in the field of view and extract the relevant photometry. At first, we run a preliminary aperture photometry procedure with a detection threshold of 10\,$\sigma$ above the local background, in order to extract only brighter sources; this selection of objects has been used as input to run a pre-PSF fitting routine, aimed at testing the optimal parameters for the PSF function to be used for analyzing the images. A final, complete single-band catalog has eventually been produced from each image processed with SExtractor by performing a new PSF-fitting run with a lower detection threshold of 3\,$\sigma$, thus recovering the fainter component of the population. 

A first filtering of each catalog population, based on the properties of the extracted sources and on the relevant extraction flags, allowed us to discard the stellar component potentially affected by saturation or other detection problems. Long and short exposure catalogs corresponding to the same spectral band, when present, have then been merged, in order to obtain a unique photometric catalog for each spectral band and avoid duplicates. The sample of objects common to both the long-exposure and short-exposure catalogs has been used to evaluate the photometric accuracy, to correct the short-exposure photometry for small magnitude offsets with photometry obtained from the long exposures and to define a confidence magnitude threshold below which photometry could begin to be affected by saturation. Objects have been preferentially taken from the long-exposure catalog; sources brighter than the long-exposure threshold have been retrieved from the short-exposure catalog, when available; sources brighter than the short-exposure threshold have been discarded.  

A complete calibration of the CFHT instrumental magnitudes to the SDSS photometric system, including a correction for both offset and color effect, has then been performed. The calibration procedure has not been anchored to a sample of standard stars in the field of view; instead, we used an ensemble of around 3000 stars, present in the CFHT FOV, for which SDSS photometry was available from the seventh SDSS Data Release \citep{SDSSDR7}, to statistically calibrate the CFHT magnitudes to the SDSS system. The matching of sources between the SDSS and the CFHT catalogs has been performed using the TOPCAT \citep{TOPCAT} tool, fixing a matching radius of \mbox{1 arcsec} and retaining only closest matches. The statistical ensemble of stars used for the calibration has been selected among the objects with complete CFHT $u^*g'r'i'$ photometry and that matched the magnitude limits for 95\% detection repeatability for point sources defined for the SDSS photometric survey \citep{SDSSDR2}. The calibration procedure adopted is based on the results of the photometric calibration study of the CFHT/MegaCam photometry to the SDSS system performed by \citet{regnault09}. We compared the distribution of SDSS colors over the ensemble of stars with the SDSS dwarf color loci tabulated in literature \citep{covey07, lenz98} and retained, for calibration purposes, only objects whose colors matched the tabulated ranges. After checking their consistency with the observed trends, Regnault's linear calibration relations have then been used to fit the ($m_\lambda^{CFHT}-m_\lambda^{SDSS}$) vs. ($m_\lambda^{SDSS}- m_{\lambda+1}^{SDSS}$) distributions in $u^*$, $g'$ and $r'$, by keeping the angular coefficient fixed and adapting the intercept value. In the $i'$-band, for which the filter in use at MegaPrime has been changed, due to accidental breakage, after the epoch of acquisition of data used in \citet{regnault09}'s study, no evident color dependence of the offset between the two photometric systems has been observed; the $i'$-band photometry has therefore been uniquely corrected for a constant offset. The system of 4 calibration equations thus obtained (one for each spectral band; Eq.\,\ref{eqn:cal_u}-\ref{eqn:cal_i}) has then been solved to obtain the final conversion from CFHT instrumental magnitudes to SDSS photometry.
\begin{equation}\label{eqn:cal_u}
u_{CFHT}-u_{SDSS}=-0.211\,\left(u_{SDSS}-g_{SDSS}\right)-0.63
\end{equation}
\vspace{-7mm}
\begin{equation}\label{eqn:cal_g}
g_{CFHT}-g_{SDSS}=-0.155\,\left(g_{SDSS}-r_{SDSS}\right)-0.51
\end{equation}
\vspace{-7mm}
\begin{equation}\label{eqn:cal_r}
r_{CFHT}-r_{SDSS}=-0.03\,\left(r_{SDSS}-i_{SDSS}\right)
\end{equation}
\vspace{-7mm}
\begin{equation}\label{eqn:cal_i}
i_{CFHT}-i_{SDSS}=-0.02
\end{equation}

A complete $u^*g'r'i'$ catalog of NGC 2264, containing $\sim$9000 sources in the field of view, has been built by cross-correlating the single-band catalogs on TOPCAT. A matching radius of \mbox{1 arcsec} has been fixed for identifying common sources detected in the $u^*,g',r'$ concurrent observations, while a larger matching radius of 2 arcsecs has been introduced to match $u^*g'r'$ sources with their $i'$-band counterpart, to account for a lower astrometric accuracy (lower angular resolution) characterizing the $i'$-band field, due to poorer seeing conditions at the time of acquisition of the $i'$-band images. Sources have been identified by retaining only best matches; we address potential problems in the derived colors associated with misidentifications and multiple matches in Sect.\,\ref{sec:catalog}. Final SDSS $u'g'r'i'$ (hereafter $ugri$) photometry has then been associated to each source in the catalog using the derived calibration relations.

\subsubsection{Monitoring survey}\label{sec:monitoring}
The second MegaCam survey of NGC~2264, performed in Feb. 2012, consisted in monitoring the $u^*$-band and $r'$-band variability of the entire stellar population of the region over a mid-term timescale (Feb.14-28, i.e., 2 weeks), as a part of the CSI~2264 project. On each observing night during the run, the region was imaged repeatedly, with a temporal cadence varying from 20\,min to 1.5\,h. Each $u^*r'$ observing block was performed using a 5-step dithering pattern, with single exposures of 3 s in the $r'$-band and 60 s in the $u^*$-band.

The procedure adopted for the photometric processing of the images obtained during the $u^*r'$ monitoring survey is similar to that applied to the mapping survey images, but each exposure has been individually processed in order to retrieve the luminosity variations over different timescales. The $i$-band image (Sect.\,\ref{sec:mapping}) has been used as the astrometric reference exposure, as the nebulosity component was the least conspicuous in this image compared to the other filters; for each source, an aperture has been placed down on each $u^*$ and $r'$ image at the position predicted from the $i$-band source detection and the amount of flux within this aperture has been extracted to build the light curve.

A preliminary quality check on the CFHT light curves consisted in closely examining a sample of NGC 2264 members with known periodicities, in order to attempt to recover their periods from the phase-folded CFHT light curves. This allowed us to ascertain the global accuracy of the $u^*r'$ time series photometry, albeit occasionally affected by isolated discrepant points and non-negligible scatter due to poorer observing conditions. In order to locate the potentially problematic observing sequences, we examined the time series photometry for the whole sample of field stars in the CFHT FOV and derived, for each observing sequence, the distribution of median zero-point offsets referred to the master frame. The zero-point offset, for a given object, is defined as the mag difference between the frame of interest and the master frame; in case of large deviation from zero, the zero-point offset distribution associated to an observing sequence thus signals that the observation has been carried out through clouds. This translates to significantly lower than average S/N for a given source, which implies that, even for the brightest stars, no accurate measurements can be inferred from that specific observation; hence, epochs matching this description have not been used for any subsequent analysis.

As done earlier for the $ugri$ snapshot survey, the instrumental light curve photometry has been calibrated to the SDSS system, referring the procedure to the calibrated photometry from the first CFHT survey. No offsets and color effects have been noticed in the $r'$-band (hereafter $r$-band) photometry, while a correction for both contributions has been applied to the $u^*$-band photometry from the ($u^*_{CFHT\_lc}-u$) vs. ($u- r$) plane:
\begin{equation}
u^*_{CFHT\_lc}-u=-0.1343\,\left(u-r\right)-0.1009
\end{equation}

\subsection{The CFHT catalog of NGC 2264} \label{sec:catalog}
Fig.\,\ref{fig:NGC2264} shows a MegaCam color picture of the field imaged at CFHT during the NGC 2264 campaign. The star-forming region extends over the central part of the field, with the most active sites of star formation located at the center of the image, northward of the Cone Nebula \citep[cf.][]{dahm08}. 

We obtained a complete $ugri$ dataset and photometric monitoring for $\sim$9000 sources in the field, exploring the area projected onto the star-forming region as well as the periphery of the cluster and background/foreground sky. Table \ref{tab:detect_limits} provides some details on the photometric properties of the population of the CFHT catalog. The survey is complete down to $u\sim$ 21.5 and $r\sim$ 18.5 and the monitored objects span a range of $\sim$7 mags. The photometry is globally accurate up to a few $\times 10^{-2}$  mag and the relative accuracy rises up to order of $10^{-3}$ mag for brighter objects.
\begin{table}
\caption{Detection limits and completeness in the CFHT $ugri$ survey of NGC 2264.}
\label{tab:detect_limits}
\centering
\begin{tabular}{c c c c c}
\hline\hline
(mag) & $u$ & $g$ & $r$ & $i$ \\
\hline
mag range\tablefootmark{*} & 15-23.5 & 14-21.5 & 13.5-20.5 & 13-19.5 \\
Saturation start & <12.5 & <13 & <13.5 & 12.5 \\
Detection limit & 23.5 & 21.5 & 20.5 & 19.5 \\
Completeness & 21.5 & 19.5 & 18.5 & 17.5 \\
\hline
\end{tabular}
\tablefoot{Values reported are conservative estimates.\\
\tablefoottext{*}{Values referred to the main body of the magnitude distribution of the catalog population.}
}
\end{table}

In order to evaluate the probability of misidentifications as a result of the procedure adopted for matching sources in different catalogs, and thus evaluate the potential impact on the photometric properties inferred for individual sources, we reconsidered all multiple identifications within the allowed matching radii and closely examined their spatial distribution. The sources with potentially multiple identifications amount to $\sim$2.9\% of the population of the catalog. The difference in offset distance between the best match and the next best match is less than 0.5 arcsec for only 23\% of these cases. Less than 20\% of the multiply matched sources are located within the region occupied by most NGC~2264 members. For the statistical purposes of this study, this component is assumed to be negligible.

\section{The PMS population of NGC 2264} \label{sec:population}
\begin{figure}
\resizebox{\hsize}{!}{\includegraphics{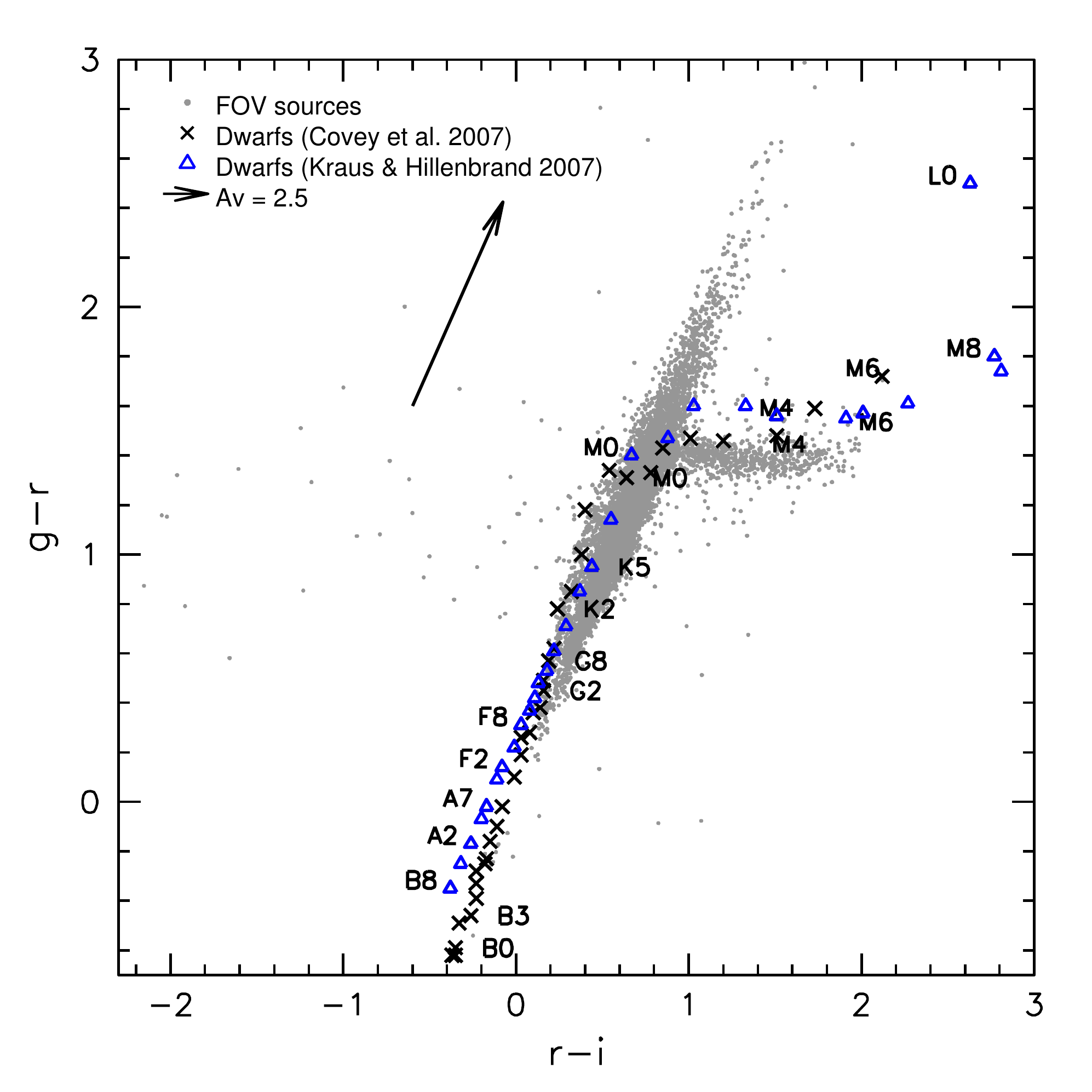}}
\caption{($r-i$, $g-r$) color-color diagram for objects monitored at CFHT. Empirical dwarf color sequences from \citet{covey07} and \citet{kraus07} are shown as black crosses and blue triangles, respectively. The reddening vector is traced based on the reddening parameters in the SDSS system provided by the Asiago Database on Photometric Systems \citep[ADPS;][]{ADPS}.}
\label{fig:ri_gr}
\end{figure}

\begin{figure}
\resizebox{\hsize}{!}{\includegraphics{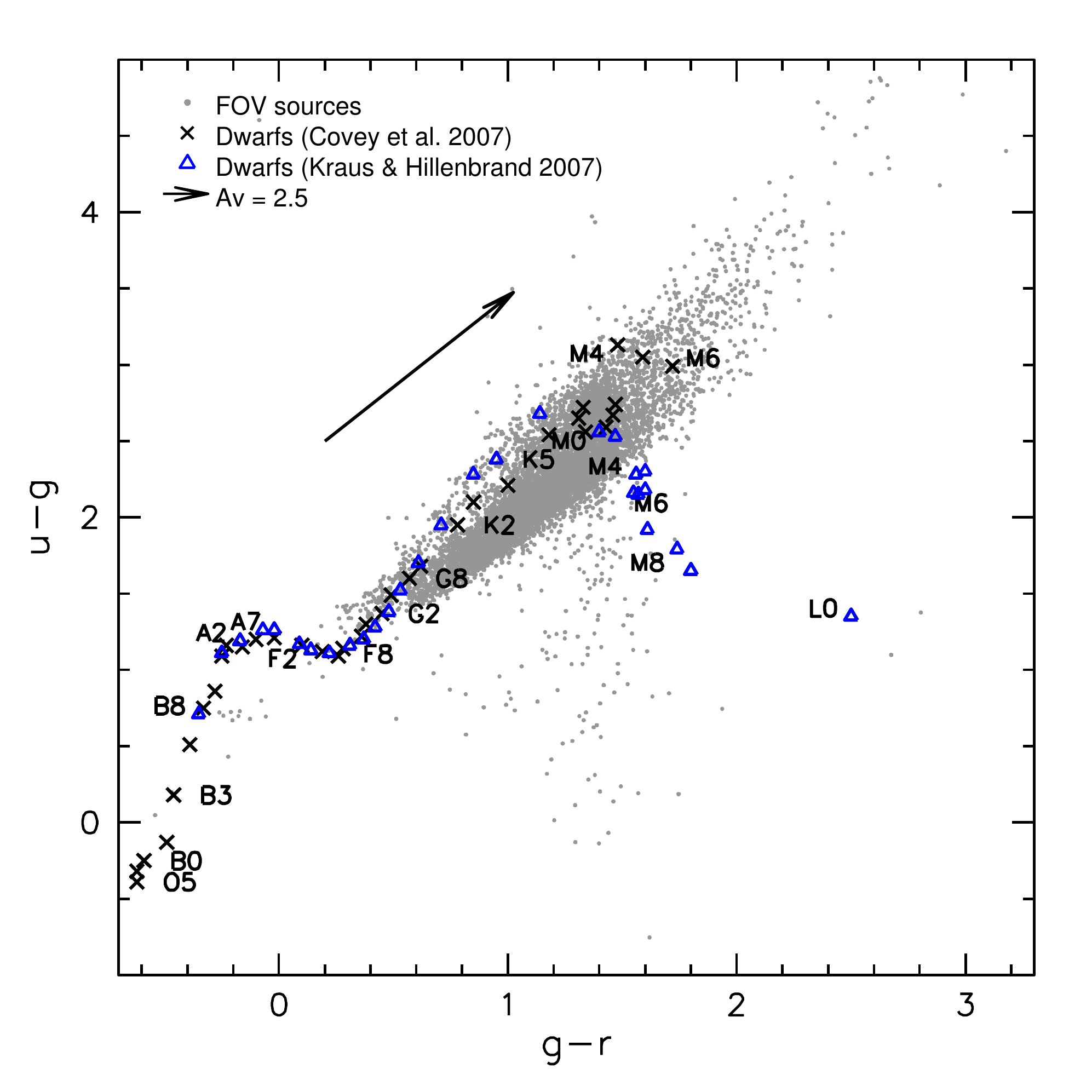}}
\caption{($g-r$, $u-g$) color-color diagram for objects monitored at CFHT. Empirical dwarf color sequences from \citet{covey07} and \citet{kraus07} are shown as black crosses and blue triangles, respectively. The reddening vector is traced based on the reddening parameters in the SDSS system provided by the ADPS \citep{ADPS}.}
\label{fig:gr_ug}
\end{figure}

A global picture of the photometric properties of the stellar population can be achieved from the analysis of color-color diagrams, as shown in Figs.\,\ref{fig:ri_gr}-\ref{fig:gr_ug}.  
Empirical dwarf SDSS color sequences \citep{covey07, kraus07} have been overplotted to the diagrams in order to follow the observed color distribution and investigate the relationship between color loci and stellar properties. 

In Fig.\,\ref{fig:ri_gr}, earlier-type field stars are located along the diagonal ellipse, with foreground or slightly reddened objects on the lowermost part of the distribution and background stars at the uppermost end; M-type stars are predominantly located on the horizontal branch, whose $g-r$ position is in good agreement with the value of $\sim$1.4 mentioned in previous studies \citep{covey07, finlator2000, F96SDSS, annis2011}. A number of objects can be located to the left (i.e., blueward) of the standard locus in $r-i$; only for a part of them this displacement may be ascribed to photometric inaccuracies. The most likely explanation for these objects is that they are highly variable, and the inferred $r-i$ color is spurious due to the large epoch difference between the $r$-band and $i$-band photometry (the over one year time lapse between the $ugr$ and the $i$-band mappings of the region, as discussed in Sect.\,\ref{sec:data}).

Similarly, a progression of increasing spectral types and reddening effects can be observed from left to right along the ellipse in Fig.\,\ref{fig:gr_ug}; an interesting locus is defined by the point dispersion at 0.5 $\leq g-r \leq$ 2.0, below the main body of the color distribution, as it shows $u-g$ color excesses presumably linked with the YSO accretion activity, manifest at UV wavelengths with a flux excess compared to the photospheric emission. An offset of a few tenths of a mag is observed between the mean locus of field stars in our survey and the sequences of \citet{covey07} and \citet{kraus07}. This offset cannot be explained with imprecisions in our calibration. A possible explanation may lie in a lower than solar standards metallicity of the field population probed here; at any rate, this disagreement has no impact on our analysis, as we uniquely used our internal reference sequence to probe the photometric accretion signatures of members, as illustrated in the following sections.

We further pursued the color signatures of different YSO types in the SDSS filters by analyzing the color properties of known members of the region with respect to the loci traced on the diagrams by the large population of field stars. Known members were identified amidst the population of the region based on a wide collection of photometric/spectroscopic data available in the literature from various surveys of the cluster; membership and further information on the WTTS (no disk evidence) vs. CTTS (disk-accreting) nature of selected members have been inferred based on one or more of the following criteria: i) H$_\alpha$ emitter from narrow-band photometry and variability from the data of \citet{lamm04} and following their criteria; ii) X-ray detection \citep{ramirez04, flaccomio06} and location on the cluster sequence in the (I, R-I) diagram when R,I photometry is available; iii) H$_\alpha$ emitter from spectroscopy (H$_\alpha$ EW > 10 \AA, H$_\alpha$ width at 10\% intensity $>$ 270 km s$^{-1}$); iv) radial velocity member according to \citet{furesz06}; v) member according to \citet{sung08} (cf. criteria enumerated earlier) and \citet{sung09} (Spitzer Class I/Class II). It is worth remarking that this preliminary analysis of membership and TTS classification is not based on direct accretion criteria. 

\begin{figure}
\resizebox{\hsize}{!}{\includegraphics{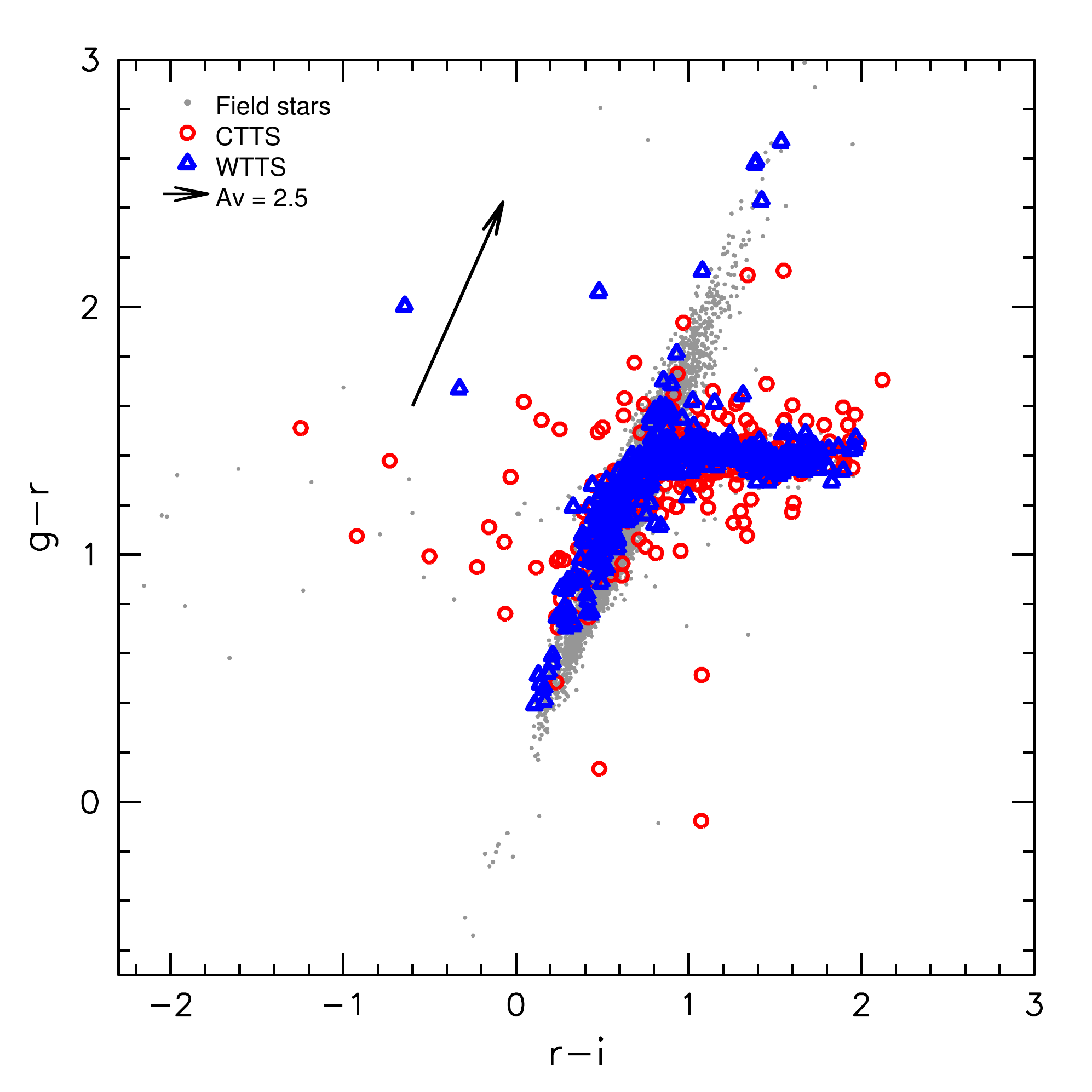}}
\caption{($r-i$, $g-r$) color-color diagram for NGC 2264 members monitored at CFHT. Field stars, non-accreting cluster members (WTTS) and accreting cluster members (CTTS) are depicted as grey dots, blue triangles and red circles, respectively. The reddening vector is traced based on the reddening parameters reported in the ADPS \citep{ADPS}.}
\label{fig:ri_gr_TTS}
\end{figure}

\begin{figure}
\resizebox{\hsize}{!}{\includegraphics{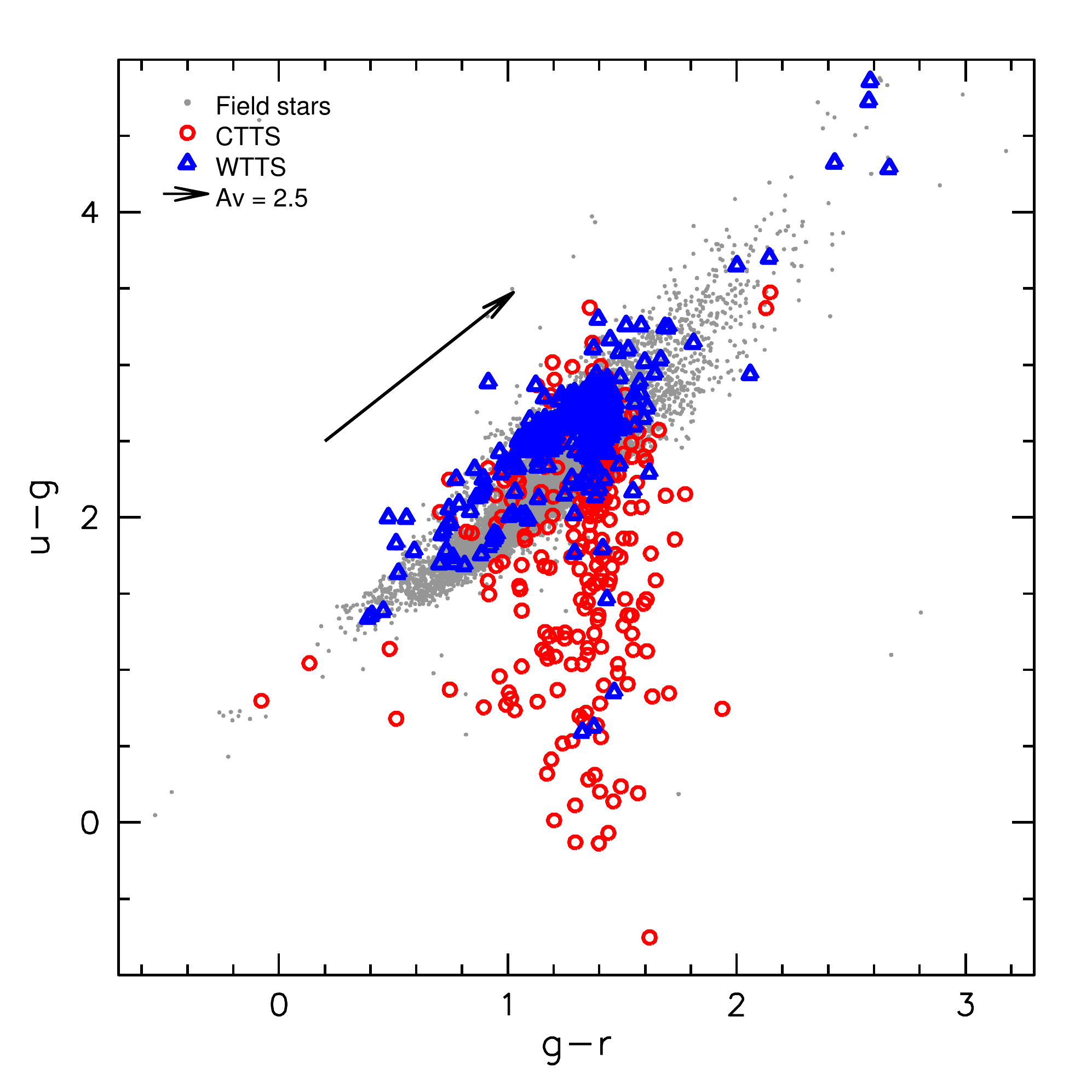}}
\caption{($g-r$, $u-g$) color-color diagram for NGC 2264 members monitored at CFHT. Field stars, WTTS and CTTS are depicted as grey dots, blue triangles and red circles, respectively. The reddening vector is traced based on the reddening parameters reported in the ADPS \citep{ADPS}.}
\label{fig:gr_ug_TTS}
\end{figure}

Around 700 known members have been matched in the CFHT catalog within a matching radius of 1 arcsec; among these, $\sim$60\% are classified as WTTS. Fig.\,\ref{fig:ri_gr_TTS} and \ref{fig:gr_ug_TTS} show the color properties of these groups of objects respectively in the ($r-i$, $g-r$) and ($g-r$, $u-g$) diagrams. 

In redder filters, as shown in Fig.\,\ref{fig:ri_gr_TTS}, both CTTS and WTTS define a single color sequence following the one traced by field stars. A few CTTS appear well blueward of this sequence, reflecting multi-year variability that is evidently more common in the CTTS than the WTTS. Conversely, accreting and non-accreting stars define two clear distinct distributions on the ($g-r$, $u-g$) diagram in Fig.\,\ref{fig:gr_ug_TTS}, with the latter displaying colors consistent with the photometric properties of field stars and the former located at smaller (i.e., bluer) $u-g$ values compared to the main color locus defined by dwarfs. 

\begin{figure}[b]
\resizebox{\hsize}{!}{\includegraphics{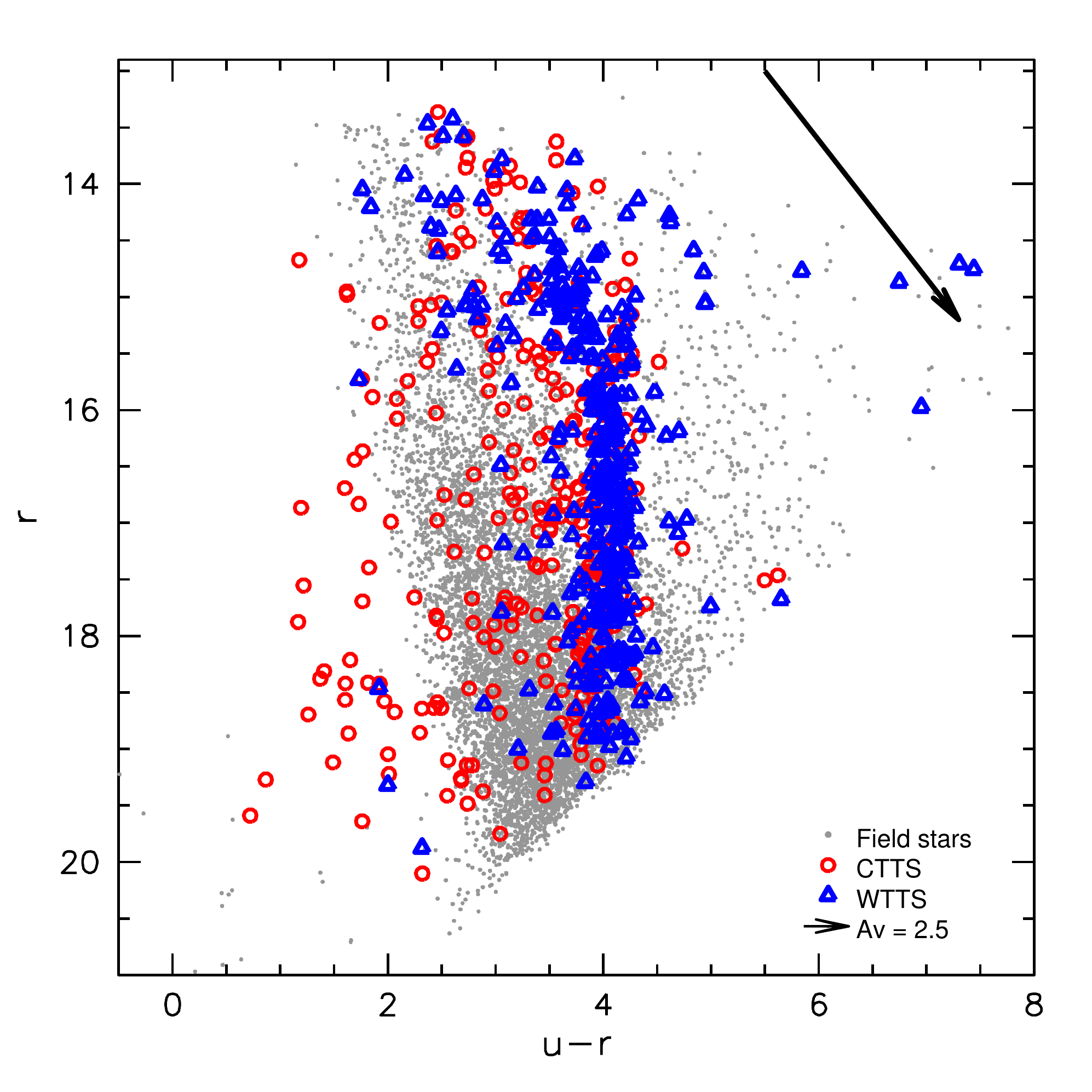}}
\caption{($u-r$, $r$) color-magnitude diagram for NGC 2264 members monitored at CFHT. Field stars, WTTS and CTTS are depicted as grey dots, blue triangles and red circles, respectively. The reddening vector is traced based on the reddening parameters reported in the ADPS \citep{ADPS}.}
\label{fig:r_ur_TTS}
\end{figure}

These distinctive features are well observed in Fig.\,\ref{fig:r_ur_TTS}, which shows how members distribute on the ($u-r$, $r$) color-magnitude diagram. There, the cluster sequence is clearly traced by WTTS, while CTTS appear broadly scattered to the left as a result of their $u$-band excess, indicative of active accretion activity, compared to WTTS. A number of presumed non-accreting members show a non-negligible displacement blueward/redward of the main locus; these two groups of objects are discussed further in Sect.\,\ref{sec:field_contam}.

Two main inferences can be drawn from these diagrams:
\begin{enumerate}
\item CTTS and WTTS members do differentiate in the SDSS colors on the basis of photometric accretion signatures at short wavelengths (UV flux excess);
\item the UV excess linked with accretion, detected in the $u$-band, does not affect significantly the observations in filters at longer wavelengths, as emphasized by the substantial agreement of the CTTS and WTTS color sequences in Fig.\,\ref{fig:ri_gr_TTS} and the color saturation at $g-r \sim$1.4, a behavior common to field stars.
\end{enumerate}
These photometric properties, observed for different ensembles of stars, enable us to perform a new membership and population study on an individual basis, from an accretion-driven perspective. This analysis has the twofold purposes of investigating the presence, in the CFHT FOV, of additional objects that are new candidate members, and of re-examining the classification of known members, looking in both cases for classification outliers in the diagram loci dominated by accretion. 

\subsection{UV excess vs. different accretion diagnostics} \label{sec:acc_diag}
In order to ascertain the coherence of the classification based on the UV excess detection on these diagrams and define some confidence limits for a reliable identification of accreting members, we compared the $u$-band excess information drawn from CFHT photometry with different diagnostics commonly used to investigate accretion, such as the equivalent width and the width at 10\% intensity in the H$\alpha$ emission line, probing the accretion funnel, or the mid-infrared emission, that signals the presence of material in the inner disk when a flux excess compared to the photospheric level is detected.

\begin{figure}[b]
\resizebox{\hsize}{!}{\includegraphics{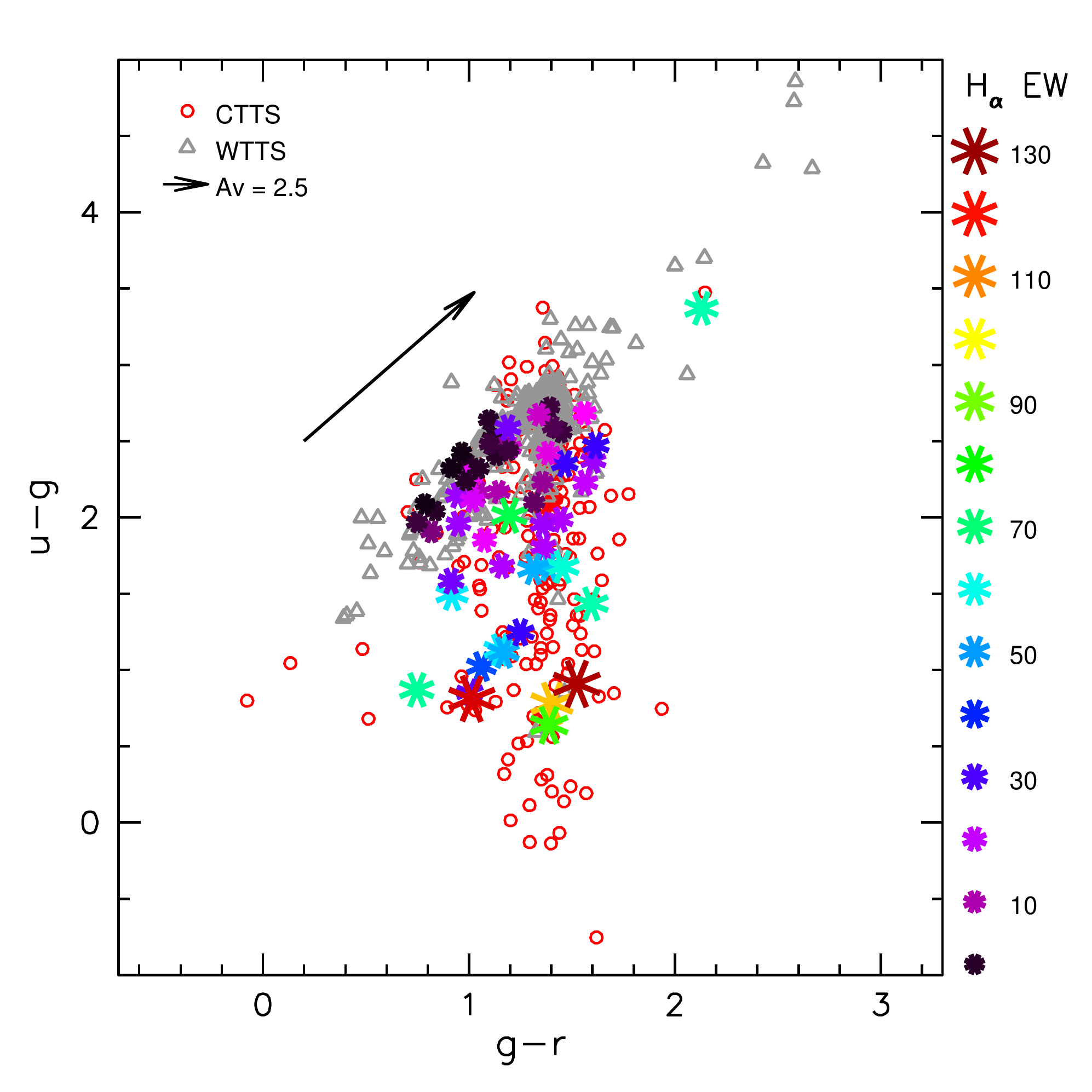}}
\caption{$u-g$ vs. $g-r$ colors for NGC~2264 members are compared to the H$_\alpha$ EW. Asterisks mark members for which H$_\alpha$ EW is available from the CSI~2264 campaign (see text). Asterisk colors and sizes are scaled according to the value of H$_\alpha$ EW.}
\label{fig:UV_HaEW}
\end{figure}

Fig.\,\ref{fig:UV_HaEW} shows the same color-color diagram as in Fig.\,\ref{fig:gr_ug_TTS} with additional information on the H$\alpha$ EW values measured for the subsample of objects having VLT/Flames spectra from the CSI~2264 campaign (A. Sousa, UFMG). A threshold value commonly adopted to identify accreting YSOs from H$\alpha$ emission is H$\alpha$ EW $\geq$ 10 $\AA$ \citep{herbig_bell1988}, while a more robust scheme relating the threshold value to the stellar spectral type has been introduced by \citet{white_basri2003}. As can be observed in Fig.\,\ref{fig:UV_HaEW}, the UV excess and the H$\alpha$ EW diagnostics are fully consistent, with the objects having the least H$\alpha$ EWs located on the color locus traced by WTTS and field stars, while a region of the diagram dominated by accretion can clearly be delimited below the field stars distribution and identified from the detection of a UV excess.
\begin{figure}[b]
\resizebox{\hsize}{!}{\includegraphics{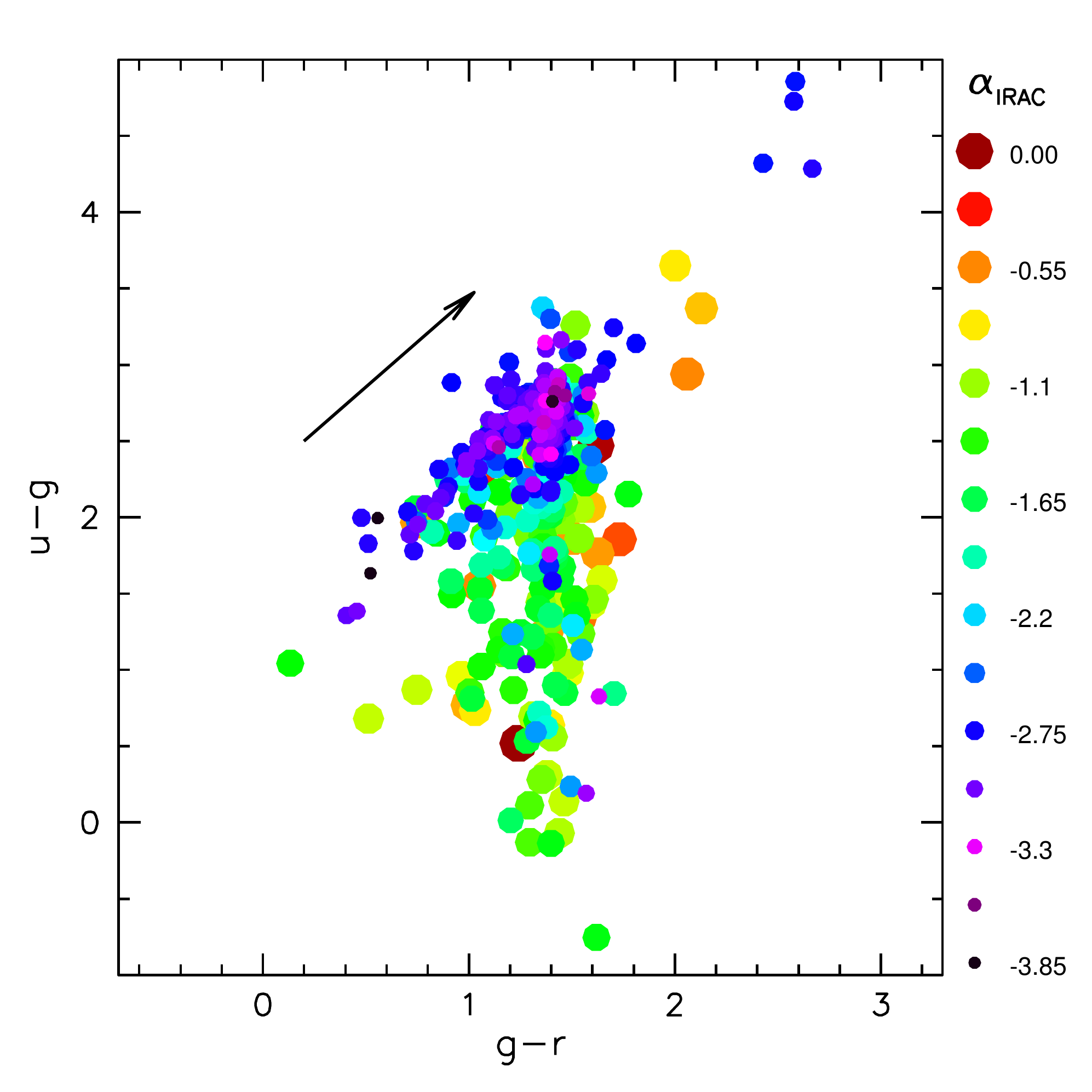}}
\caption{$u-g$ vs. $g-r$ colors for NGC~2264 members are compared to the $\alpha_{IRAC}$ diagnostics, which probes the inner disk properties (see text). Colors and sizes of symbols are scaled according to the value of $\alpha_{IRAC}$ from the data of \citet{teixeira2012}.}
\label{fig:UV_alpha_irac}
\end{figure}

A useful tool for characterizing presence and properties of a circumstellar disk is the examination of the infrared spectral energy distribution (SED) of the system. A particularly interesting indicator of the SED morphology is the measure of its slope in the spectral range of interest \citep[i.e., the spectral index $\alpha$ first introduced by][]{lada1987}: smaller amounts of material in the inner disk result in smaller contributions, i.e. smaller flux excesses, at mid-IR wavelengths and hence more negative values of the $\alpha$-index, which approaches the slope of a reddened blackbody, while positive values of the mid-IR SED slope are specific to deeply embedded, Class I sources. In \citet{lada06alphaIRAC}, the $\alpha$-index diagnostics for disks is applied to the mid-IR emission sampling between 3.6 and 8 $\mu$m performed with the Spitzer instrument IRAC for the young cluster IC~348; the same tool has been adopted by \citet{teixeira2012} to characterize the evolutionary state of NGC~2264 members based on the inner disk properties. We compare their results to the colors of NGC~2264 YSOs in the ($g-r$, $u-g$) diagram in Fig.\,\ref{fig:UV_alpha_irac}. Following the classification proposed in Table 2 of \citet{teixeira2012}, values of $\alpha_{IRAC}\la-2$ are indicative of objects with anaemic disks or naked photospheres, while objects with thick disks (hence likely to be actively accreting) are characterized by $\alpha_{IRAC}>-2$. Indeed, as can be observed in Fig.\,\ref{fig:UV_alpha_irac}, objects in the first $\alpha_{IRAC}$ group are mainly located on the WTTS/dwarf locus in $u-g$ vs. $g-r$, while objects showing a UV excess (and thus actively accreting stars) have the largest values of $\alpha_{IRAC}$, as globally expected. 

\subsection{UV census of the NGC~2264 population} \label{sec:census}
\subsubsection{New CTTS candidates in NGC 2264} \label{sec:cand}
\begin{figure}
\resizebox{\hsize}{!}{\includegraphics{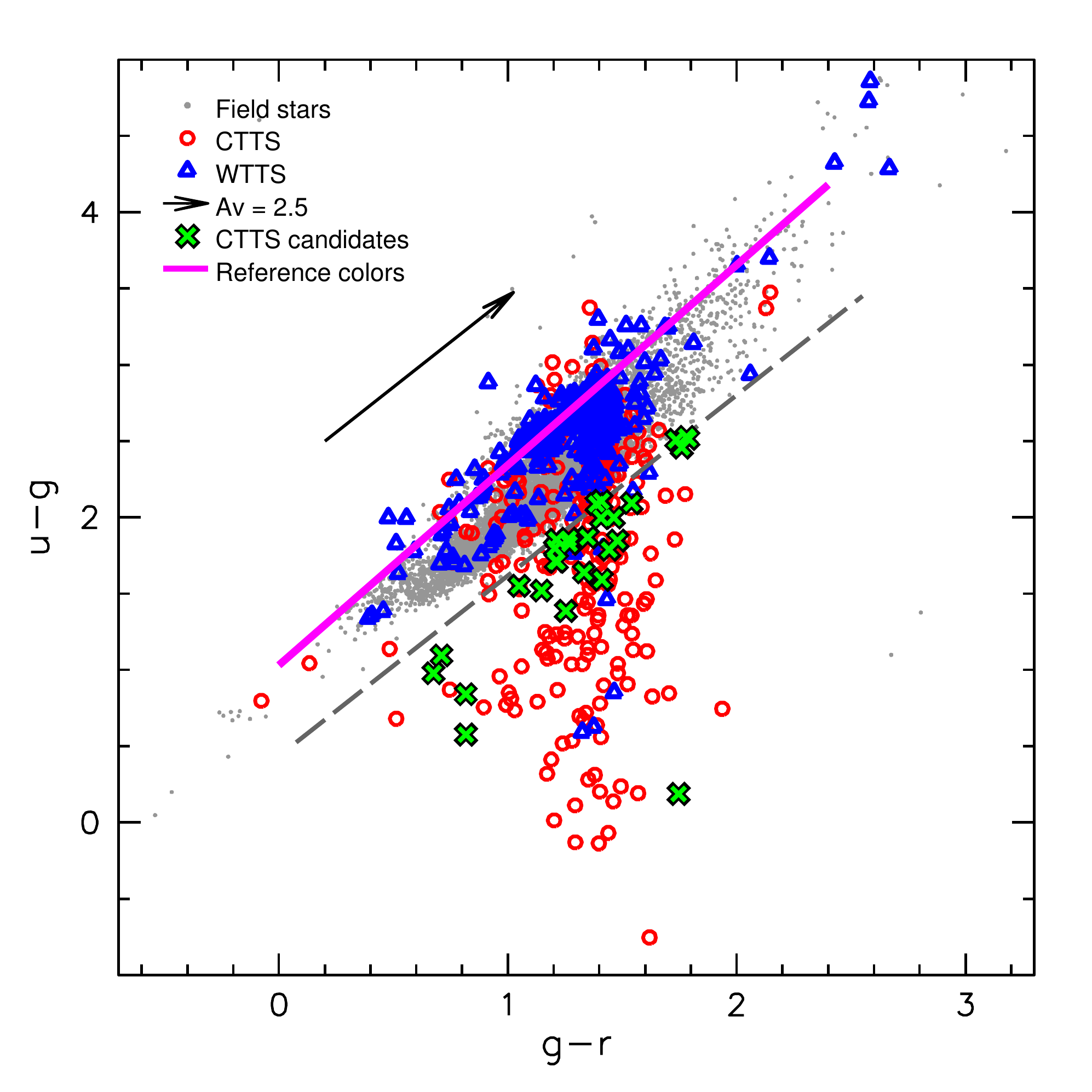}}
\caption{Selection of new CTTS candidates from the ($g-r$, $u-g$) diagram. The grey dashed line traces a conservative boundary separating the accretion-dominated region of the diagram from the color locus of non-accreting stars (see text). Field stars, WTTS and known CTTS are depicted as grey dots, blue triangles and red circles, respectively; the new CTTS candidates selected from this diagram are marked as green crosses. The purple line fitting the upper envelope of the WTTS distribution traces the reference sequence of photospheric colors that has been adopted for measuring the UV excess of accreting stars (see Sect.\,\ref{sec:accretion}).}
\label{fig:gr_ug_cand}
\end{figure}
A most interesting locus to identify accreting members is the data points dispersion below the main body of the distribution in Fig.\,\ref{fig:gr_ug_TTS}. In fact, this allows easy detection and measurement of the $u$-band excess with respect to WTTS. The inferred color excess estimate is independent of distance and has no significant dependence on visual extinction A$_V$, as WTTS define a color sequence nearly parallel to the reddening vector down to $\sim$M0 spectral type. For later-type stars, the color sequence traced by WTTS turns down, saturating at $g-r$\,$\sim$\,1.4 (cf.~Fig.\,\ref{fig:ri_gr_TTS}); the implications of this for the UV excess measurement are discussed in Sect.\,\ref{sec:accretion}. In order to minimize the contamination, we traced a straight line, parallel to the reddening vector, bordering the accretion-dominated area below the lower envelope of the dwarf color locus (Fig.\,\ref{fig:gr_ug_cand}); we then selected as new CTTS candidates all the objects lying in this region and marked as field stars. For all these objects, a preliminary check of the photometry on the CFHT field images has been performed, in order to discard the sources whose flux measurements could be affected by detection issues (e.g., too faint source, presence of two close sources, partial projection of the source on CCD gaps or bad pixel rows). 

A second group of new CTTS candidates has been extracted from the color distribution in Fig.\,\ref{fig:ri_gr_TTS}, namely among the color outliers located to the left and, to a smaller extent, below the main color distribution. Similarly to the procedure adopted for the selection of objects on the diagram in Fig.\,\ref{fig:gr_ug_TTS}, a check on the photometry quality for all the objects of interest has preceded the identification of new candidates. 

\begin{figure}
\resizebox{\hsize}{!}{\includegraphics{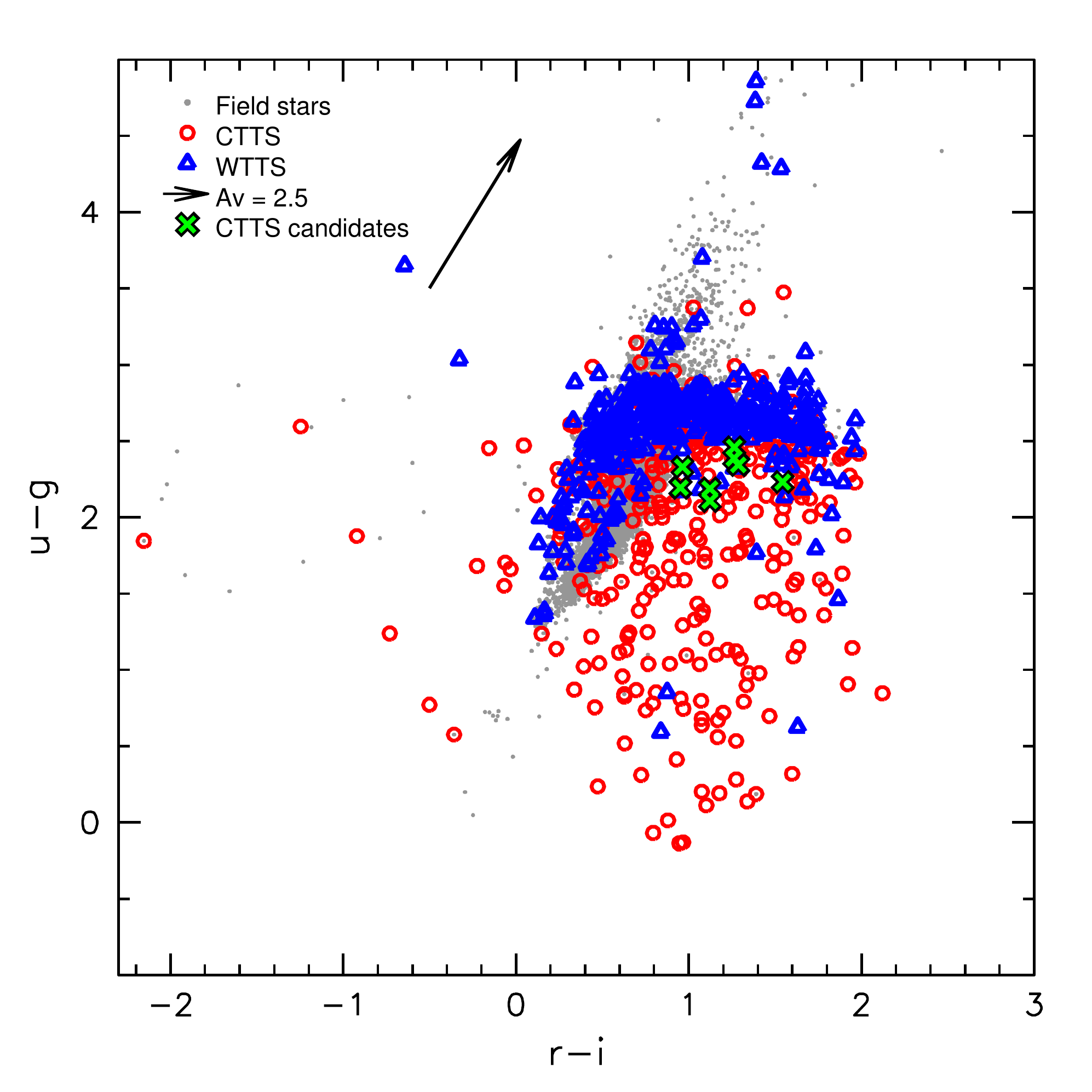}}
\caption{($r-i$, $u-g$) diagram for cluster members and field stars in the NGC~2264 field. Field stars, WTTS and known CTTS are depicted as grey dots, blue triangles and red circles, respectively. Grey dots encircled in red correspond to new CTTS candidates selected on different diagrams (Fig.\,\ref{fig:gr_ug_cand} or Fig.\,\ref{fig:ri_gr_TTS}.) Green crosses mark the additional CTTS candidates selected on this diagram.}
\label{fig:ri_ug_cand}
\end{figure}

A third color-color diagram of interest for the analysis of membership is the one depicting $u-g$ vs. $r-i$ (Fig.\,\ref{fig:ri_ug_cand}); as a ``transitional'' diagram between $u-g$ vs. $g-r$ (Fig.\,\ref{fig:gr_ug_TTS}; direct UV excess detection) and $g-r$ vs. $r-i$ (Fig.\,\ref{fig:ri_gr_TTS}; nearly horizontal photospheric color sequence for late-type stars, with no color saturation on $r-i$), this diagram provides a greater sensitivity to the UV excess for spectral types $\geq$M0. Additional, later-type CTTS candidates have then been identified on this diagram, and their photometric properties inspected similarly to what done for the candidate members extracted earlier.

A second aspect of our selection of new candidate, actively accreting YSOs concerns the group of already known members which had previously been considered WTTS (that is, YSOs without evidence of significant, ongoing accretion at the time of the previous surveys). We examined the locations of these putative WTTS in our several diagrams (Figures \ref{fig:ri_gr_TTS} through \ref{fig:gr_ug_cand}) and identified a significant number whose colors fell generally within the UV-bright regions -- for example, the small group of blue points located below the normal dwarf locus in Fig.\,\ref{fig:gr_ug_TTS}. This procedure allowed us to re-classify 19 WTTS members as CTTS, which add to the 50 newly accretion-identified candidates; this varied status does not necessarily reflect an erroneous previous classification, but may as well be the result of a long-term ($\sim$years) variability in the accretion activity of the objects.

The results of our UV-based membership investigation are provided in Table\,\ref{tab:cand} (new CTTS candidates) and Table\,\ref{tab:reclass} (re-classified CTTS). This revised census of accreting members we infer for NGC~2264 likely encompasses all objects showing a strong UV excess, hence a significant amount of accretion. However, our sample might not be exhaustive regarding the weakly-accreting CTTS component. Indeed, a non-negligible overlap between the CTTS and WTTS populations can be seen in Figs.\,\ref{fig:gr_ug_TTS}-\ref{fig:r_ur_TTS}, which cannot be efficiently probed with our diagnostics. Hence, we do identify some new CTTS, but we might be missing a number of lower accretors, showing a smaller UV excess that cannot be easily separated from the color properties of WTTS/field stars. 

\begin{figure}
\centering
\includegraphics[width=8cm]{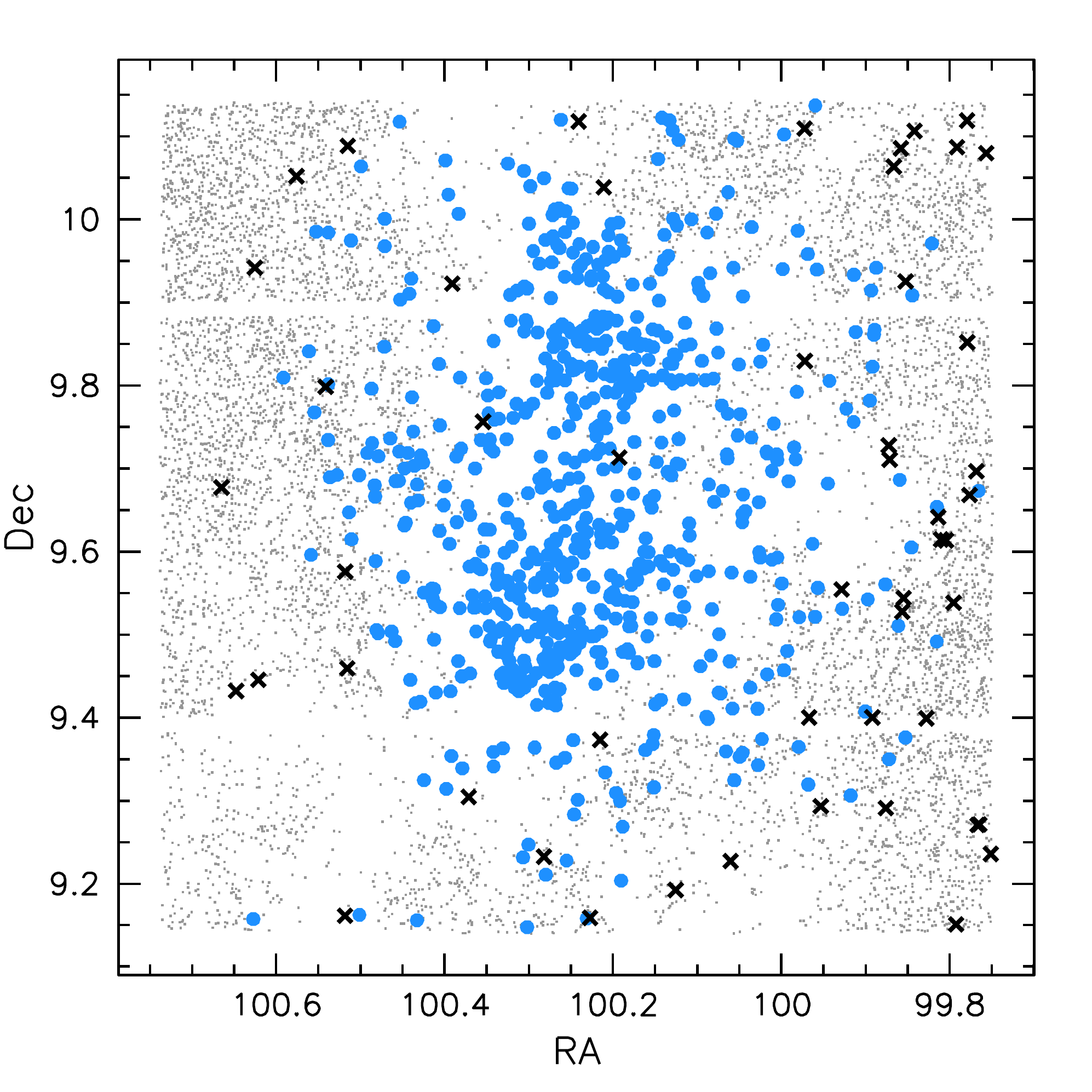}
\caption{Spatial distribution of the new CTTS candidates from the CFHT $u$-band survey. Field stars, known cluster members and new CTTS candidates are depicted as small grey dots, blue dots and black crosses, respectively.}
\label{fig:cand_ra_dec}
\end{figure}

The wide FOV of CFHT/MegaCam allowed us to probe a quite extended region around the cluster, encompassing most of the fields of view adopted in previous surveys. It is thus of interest to examine the spatial distribution of the new CTTS candidates, selected as described in Sect.\,\ref{sec:cand}, compared to the distribution of known NGC~2264 members on the CFHT field of view. This comparison is shown in Fig.\,\ref{fig:cand_ra_dec}. A fraction of the object positions are projected onto the main cluster region, while a significant part are located on the periphery or beyond. This provides evidence for a wider spatial spread of the PMS population associated with NGC~2264, and attests to the presence of actively accreting members far from the original sites of star formation, currently known to be located towards the north and the center of the field imaged at CFHT \citep{dahm08}. 

\subsubsection{Field stars contamination in the NGC~2264 sample}\label{sec:field_contam}
Based on the discussion in Sect.\,\ref{sec:cand}, another potentially interesting location for our accreting members census is the data point dispersion to the left of the cluster sequence (WTTS distribution) in Fig.\,\ref{fig:r_ur_TTS}. A direct search for candidate members in this color-magnitude locus would hardly be  straightforward, due to the distance--dependence and to the vast preponderance of field stars; on the other hand, this diagram provides an additional test bench for investigating the nature of individual groups of objects. 

Similarly to Fig.\,\ref{fig:gr_ug_TTS}, we can identify on Fig.\,\ref{fig:r_ur_TTS} a group of putative non-accreting members with inferred $u-r$ colors bluer than expected. This feature, suggestive of ongoing accretion activity for a young cluster member, is consistently observed on different color-magnitude diagrams ($g$ vs. $u-g$, $i$ vs. $u-i$) for several of these objects; interestingly, these objects populate the thin lower branch of the WTTS distribution, below the main WTTS locus, on the lower part of the dwarf locus in Fig.\,\ref{fig:gr_ug_TTS} ($u-g$ vs. $g-r$ diagram). Little spectroscopic information is available for this group of stars; contrary to what is expected for objects clearly showing accretion activity, the level of variability exhibited by many of these objects during our monitoring is not significant over the noise level defined, as a function of magnitude, by field stars in our survey (see Venuti et al., in prep.). Additionally, several of these putative members would be located close to or onto the MS-turnoff (or below) in the HR diagram of the region (Fig.\,\ref{fig:HRD}). We therefore suspect at least a part of these objects to possibly be misclassified dwarfs instead. A list of objects with questioned membership, of the relevant CFHT photometry and of the original selection criteria is provided in Table\,\ref{tab:questioned}. 

Another interesting group of objects in Fig.\,\ref{fig:r_ur_TTS} is the sample of WTTS displaced redward of the main cluster sequence. Most of these objects consistently appear significantly reddened through all CFHT photometric diagrams; in several cases, they display little variability. While contamination by field stars might be partly reflected in the observed properties, most of the objects in this stellar group would actually look like very young, or equivalently overluminous, objects on the HR diagram of the region (i.e., they are located on the upper right envelope of the data point distribution in Fig.\,\ref{fig:HRD}). This would rather suggest binarity and/or earlier (i.e., more embedded) evolutionary status as possible explications of the distinctive properties of these objects.

\subsection{NGC 2264 members in the CFHT survey: individual extinction and stellar parameters}\label{sec:parameters}
The sample of PMS stars retrieved or newly identified in the CFHT survey amounts to about 750 objects. Based on the multiband photometry obtained during our campaign, as well as on the collection of previous literature data in different spectral bands, we performed an extensive investigation of individual stellar parameters (like extinction, mass and radius); this is aimed at characterizing the properties of members in a uniform and self-consistent manner.

Extinction towards NGC 2264 is known to be low \citep[cf.][for a review of literature results]{dahm08}, with a typical value of $\la$\,0.4 mag for the visual extinction A$_V$. Indeed, the good agreement of the cluster sequence with the color locus traced by foreground stars on Fig.\,\ref{fig:ri_gr_TTS} and the color saturation at $\sim$1.4 in $g-r$ (cf. discussion at the beginning of Sect. \ref{sec:population}) is consistent with a low value of the average A$_V$. We use the photometric properties displayed in this diagram to derive an estimate of the individual A$_V$ for late-type stars, i.e. the group of members located on the horizontal branch in the color locus. For this group of objects, A$_V$ is determined by displacing each point along the reddening direction until the lowermost envelope of the branch distribution defined by field stars is reached and computing the corresponding amount of reddening. This procedure has the advantage of relying solely on observed quantities, without involving the comparison with an external reference sequence that would intrinsically introduce an additional source of uncertainty on the result. 

For earlier-type stars ($\la$M0), the cluster sequence is nearly parallel to the reddening direction in Fig.\,\ref{fig:ri_gr_TTS}, thus preventing a direct A$_V$ determination from the object location on the diagram. For this group of objects, we derive a first estimate of the individual A$_V$ from the analysis of the infrared photometry in the JHK$_S$ filters, as measured in the 2MASS survey \citep{2MASS}. This consists in comparing their infrared colors with the JHK$_S$ color sequence for dwarfs \citep{covey07} and with the CTTS infrared locus defined in \citet{meyer1997} and converted to the 2MASS photometric system as in \citet{covey2010} (K. Covey, private communication). Additional A$_V$ estimates for a part of these sources are available from previous, similar studies (e.g., \citealp{rebull2002}, where A$_V$ is investigated from the color excess on R-I, or \citealp{cauley2012}, where A$_V$ is computed by fitting stellar spectra with spectral templates) and/or derived from our optical photometry collection (from the measurement of the color excess on R-I, V-I or B-V in order of preference). In the absence of a direct A$_V$ estimate from CFHT photometry, all available A$_V$ derivations have been examined and their consistency checked on the color properties displayed in the CFHT diagrams, in order to discard discrepant estimates; an average value has then been adopted as a final estimate. 

In order to evaluate the uncertainty on our individual A$_V$ estimates, we collected all available A$_V$ determinations for any object in our sample, and measured the amount of scatter observed among values inferred from different authors/methods. This comparison allowed us to conclude that different A$_V$ estimates are typically consistent within a radius of a few 0.1 mag, which then provides an order-of-magnitude uncertainty on the A$_V$ values adopted in this study.

Of the sample of CFHT members, spectral type is known for around 50\% and has been retrieved, in order of preference, from the studies of \citet{dahm05}, \citet{rebull2002} or \citet{walker1956}. For the remaining objects, the spectral type has been derived from the comparison of the dereddened colors with the empirical optical color sequence of \citet{covey07}. We used the dereddened photometry of members with known spectral type, earlier than M1, to recalibrate the empirical spectral type--color sequence onto CFHT photometry. Spectral types were converted to effective temperatures T$_{eff}$ following the scale of \citet{cohen1979}, checked for consistency against the scale of \citet{luhman2003}.

Bolometric luminosities have been derived from the dereddened J-band photometry retrieved from the 2MASS catalog, adopting T$_{eff}$--dependent J-band bolometric corrections (BC$_J$) obtained as a fit to the T$_{eff}$--BC$_J$ scales of \citet{pecaut2013} and \citet{bessell98}. A cluster distance value of 760~pc \citep{sung97} has been adopted for the conversion to absolute magnitudes; this commonly adopted estimate of distance has been recently strengthened by the results of \citet{gillen2014}, who performed a detailed characterization of a newly identified, well sampled low-mass PMS eclipsing binary in NGC~2264, with strong evidence of membership and an inferred distance of 756$\pm$96~pc. For a few tens of objects, for which J-band photometry from the 2MASS survey was not available, L$_{bol}$ has been derived using an empirical calibration relationship between the dereddened $r$-band photometry and L$_{bol}$, which has been inferred from the distribution of dereddened $r$-band magnitudes vs. L$_{bol}$ values derived from the J-band photometry over the whole sample.

\begin{figure}
\resizebox{\hsize}{!}{\includegraphics{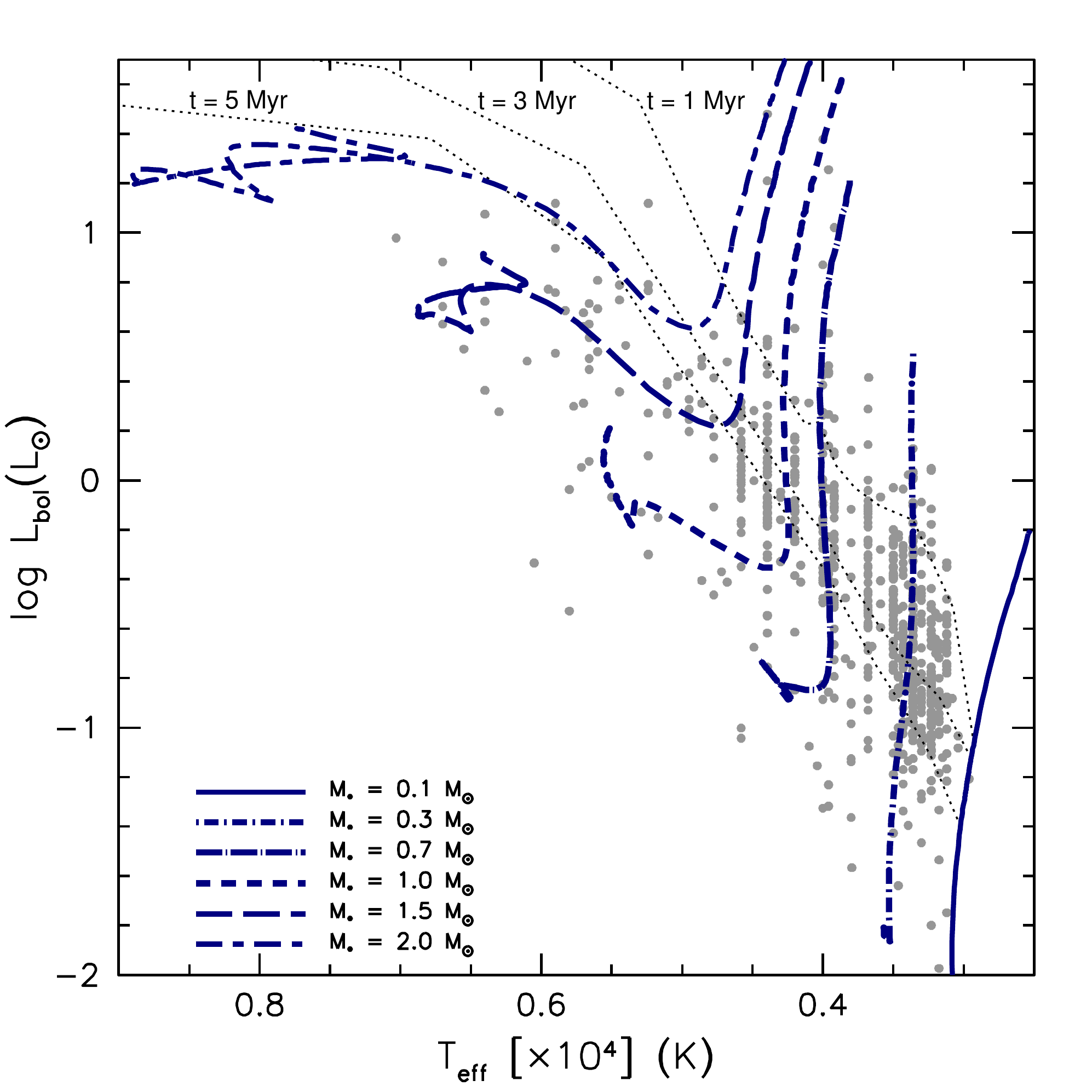}}
\caption{HR diagram for NGC 2264 members. Mass tracks and isochrones from \citet{siess2000}'s models are plotted over the point distribution.}
\label{fig:HRD}
\end{figure}

Stellar masses have been determined by placing each object on the Hertzsprung-Russell diagram and comparing their position with the mass tracks of the PMS model grid of \citet{siess2000} (Fig.\,\ref{fig:HRD}). A track-fitting tool provided by L. Siess on his webpage has been used to interpolate the mass value between two mass tracks. CFHT members span a wide range of masses, varying from 0.1 to 2 M$_\odot$. 
Stellar radii have been determined from the estimates of L$_{bol}$ and T$_{eff}$ as
$$
R_{*}=\sqrt{\frac{L_{bol}}{4\pi\sigma T_{eff}^4}}
$$
where $\sigma$ is the Stefan-Boltzmann constant.

Photometric data and stellar parameters for the whole population of NGC~2264 members monitored at CFHT are reported in Table \ref{tab:phot} and Table \ref{tab:param}, respectively.

\begin{table*}
\caption{Photometry for NGC~2264 members monitored at CFHT.}
\label{tab:phot}
\centering
\begin{tabular}{c c c c c c c c}
\hline\hline
  \multicolumn{1}{c}{Mon-ID\tablefootmark{1}} &
  \multicolumn{1}{c}{RA} &
  \multicolumn{1}{c}{Dec} &
  \multicolumn{1}{c}{$u$} &
  \multicolumn{1}{c}{$g$} &
  \multicolumn{1}{c}{$r$} &
  \multicolumn{1}{c}{$i$} \\
\hline
  Mon-000007 & 100.47109 & 9.96755 & 15.850 & 14.807 & 14.674 & 14.192 \\
  Mon-000008 & 100.45248 & 9.90322 & 19.656 & 17.083 & 15.839 & 19.887 \\
  Mon-000009 & 100.53812 & 9.80134 & 18.208 & 16.043 & 15.011 & 14.532 \\
  Mon-000011 & 100.32187 & 9.90900 & 18.548 & 16.808 & 15.531 & 14.535 \\
  Mon-000014 & 100.52772 & 9.69214 & 18.895 & 15.879 & 14.280 & 13.445 \\
  Mon-000015 & 100.53796 & 9.98410 & 19.427 & 16.172 & 14.589 & 13.786 \\
  Mon-000017 & 100.38330 & 10.00680 & 18.402 & 15.972 & 14.930 & 14.496 \\
  Mon-000018 & 100.30520 & 9.91909 & 19.311 & 16.526 & 15.366 & 14.613 \\
 Mon-000020 & 100.53849 & 9.73427 & 19.964 & 17.213 & 15.891 & 15.267 \\
  Mon-000021 & 100.24772 & 9.99594 & 18.174 & 15.835 & 14.810 & 14.452 \\
\hline
\end{tabular}
\tablefoot{A full version of the Table is available in electronic form at the CDS. A portion is shown here for guidance regarding its form and content. Uncertainties on the photometric measurements are estimated to range from $\sim$0.15 to $\sim$0.004 from the faintest to the brightest objects in $u$ and $g$, from $\sim$0.015 to $\sim$0.0015 in $r$ and 5-10 times smaller in $i$.\\
\tablefoottext{1}{Object identifiers adopted within the CSI~2264 project \citep[see][]{cody2014}.} 
}
\end{table*}

\begin{table*}
\caption{Spectral type, extinction and stellar parameters for NGC~2264 members monitored at CFHT.}
\label{tab:param}
\centering
\begin{tabular}{c c c c c c c c c}
\hline\hline
  \multicolumn{1}{c}{Mon-ID} &
  \multicolumn{1}{c}{Status\tablefootmark{1}} &
  \multicolumn{1}{c}{SpT} &
  \multicolumn{1}{c}{SpT\_ref.\tablefootmark{2}} &
  \multicolumn{1}{c}{A$_V$} &
  \multicolumn{1}{c}{L$_{bol}$(L$_\odot$)} &
  \multicolumn{1}{c}{M$_*$(M$_\odot$)} &
  \multicolumn{1}{c}{R$_*$(R$_\odot$)} &
  \multicolumn{1}{c}{$\log$\,t(yr)}\\
\hline
  Mon-000007 & c & K7 & s &  & 1.28 & 0.69 & 2.36 & 6.09\\
  Mon-000008 & w & K5 & s & 0.3 & 0.241 & 0.80 & 0.85 & \\
  Mon-000009 & w & F5 & p & 2.3 & 2.31 & 1.3 & 1.24 & \\
  Mon-000011 & c & K7 & s & 0.3 & 0.79 & 0.70 & 1.86 & 6.34\\
  Mon-000014 & w & K7:M0 & p & 1.2 & 3.9 & 0.66 & 4.22 & 5.68\\
  Mon-000015 & w & K7:M0 & p & 1.0 & 2.70 & 0.65 & 3.50 & \\
  Mon-000017 & c & K5 & s & 0.2 & 0.71 & 1.13 & 1.45 & \\
  Mon-000018 & w & K3:K4.5 & p & 0.6 & 1.70 & 1.47 & 2.03 & 6.60\\
 Mon-000020 & w & K7 & s & 0.4 & 0.44 & 0.71 & 1.39 & 6.70\\
  Mon-000021 & c & K5 & s & 0.4 & 0.96 & 1.20 & 1.69 & 6.69\\
\hline
\end{tabular}
\tablefoot{A full version of the Table is available in electronic form at the CDS. A portion is shown here for guidance regarding its form and content.\\
\tablefoottext{1}{``c'' = CTTS; ``w'' = WTTS; ``cc'' = CTTS candidate.}\\
\tablefoottext{2}{``s'' = spectroscopic SpT estimate (retrieved from literature); ``p'' = photometric SpT estimate (from CFHT optical colors).}
}
\end{table*}

\section{UV excess and accretion in NGC 2264} \label{sec:accretion}
The investigation of YSOs at short wavelengths offers a unique window on the accretion mechanisms: it provides one of the most direct probes to the hot emission from the impact layer where the accreting material hits the stellar surface at near free-fall velocities, producing a shocked area up to several $\times$\mbox{$10^3$ K} hotter than the stellar photosphere. This enhanced luminosity at UV wavelengths is reflected in the characteristic color properties of accretion-dominated objects compared to non-accreting young stars (cf. Figs.\,\ref{fig:gr_ug_TTS} and \ref{fig:r_ur_TTS}) and hence provides a direct proxy to the accretion luminosity, which in turn enables the investigation of the rates of mass accretion in individual systems. 

The $u$-band flux excess is trivially given by
\begin{equation}\label{eqn:f_exc}
F_u^{exc}=F_u^{obs}-F_u^{phot},
\end{equation}
where $F_u^{obs}$ is the measured flux in the $u$-band and $F_u^{phot}$ is the amount of flux that would be expected from a purely photospheric emission. It is important to remark that, in order to isolate the component of additional flux produced in the accretion shock and thus probe accretion onto the star, the flux excess (compared to dwarfs) due to the enhanced chromospheric activity of T Tauri stars has to be included in the photospheric emission. This issue is addressed by taking the photometric properties observed for the WTTS population as the reference for the definition of the UV excess. The color excess will thus correspond to the difference between the observed stellar colors and the expected photospheric colors based on the properties of the non-accreting counterparts of the objects of interest:
\begin{equation}\label{eqn:uv_exc}
E(u-m) = (u-m)_{obs} - (u-m)_{phot+chrom}
\end{equation}
As discussed in Sect. \ref{sec:population}, no significant impact of the accretion luminosity is observed on the stellar flux detected at filters redder than the $u$-band; hence, the color excess of Eq.\,\ref{eqn:uv_exc} basically corresponds to the $u$-band excess revealing the presence of accretion, $E(u) \simeq E(u-m)$.

We measured the UV color excess of accreting stars in two different ways, from the properties displayed on the $u-g$ vs. $g-r$ and $r$ vs. $u-r$ diagrams (Fig.\,\ref{fig:gr_ug_TTS} and Fig.\,\ref{fig:r_ur_TTS}, respectively), as detailed below. 
\begin{itemize}
\item In the first case (Fig.\,\ref{fig:gr_ug_TTS}), we referred the color excess measurement to a straight line nearly parallel to the reddening vector and following the upper part of the WTTS distribution, as shown in Fig.\,\ref{fig:gr_ug_cand}. $E(u)$ is thus measured as
\begin{equation}\label{eqn:uv_exc_ug}
E(u) = (u-g)_{obs} - (u-g)_{ref}\,,
\end{equation}
where $(u-g)_{ref}$ is the reference (i.e., expected) color at the observed $g-r$ level, $(u-g)_{ref}=1.312\,(g-r)_{obs}+1.03$. The WTTS color sequence is in good agreement with the reference line until spectral type $\sim$M0, and hence the adoption of this reference sequence allows us to derive a UV excess estimate essentially unaffected by reddening. For later-type stars, however, the real WTTS color sequence turns down on the diagram; this implies that the extrapolation of the WTTS trend at spectral types earlier than M0 to the whole sample yields an excess overestimation (i.e., more negative values of $E(u)$) for M-type stars. As the bulk of M-type WTTS lies about 0.2 mag below the reference line, we corrected for this bias by adding a constant offset of 0.2 mag to the $E(u)$ estimates derived for this group of objects. \\
\item In the second case, as can be observed on Fig.\,\ref{fig:r_ur_TTS}, WTTS nicely trace the cluster sequence over the whole range of magnitude. We thus dereddened the $u,r$ photometry and defined the sequence of reference colors as the least-squares fit polynomial to the resulting WTTS distribution. The UV excess is thus provided by
\begin{equation}\label{eqn:uv_exc_ur}
E(u) = (u-r)_{obs} - (u-r)_{ref}\,,
\end{equation}
where $(u-r)_{ref}$ is the reference color at brightness $r_{obs}$. Due to the small number of point and the large scatter affecting the definition of the sequence at the brighter end, we decided to restrict our UV excess analysis to objects fainter than 14.5 in $r$.

This second method, compared to the first, has the disadvantage of being intrinsically more uncertain, as it is subject to errors in A$_V$ determination (while the first method is essentially A$_V$-independent); on the other hand, the application of this method to $u$- and $r$-band observations taken during the monitoring survey allows us to derive both a global picture of accretion in NGC 2264 from the average photometry, and a variability range for the UV excess on a timescale of a few weeks (cf. Sect.\,\ref{sec:macc_var}). 
\end{itemize}
An average rms error of $\sim$0.16 mag has been associated to the UV excess determinations, in order to account for the scatter of the WTTS distribution around the reference sequence.
\\

From Eq.\,\ref{eqn:f_exc} and
\begin{equation}\label{eqn:obs_phot}
E(u) = u_{obs} - u_{phot}
\end{equation}
we obtain
\begin{equation}\label{eqn:fu_exc}
F_u^{exc}=F_u^{obs} \left(1-10^{+0.4\,E(u)}\right)=F_u^0\,10^{-0.4\,u}\left(1-10^{+0.4\,E(u)}\right),
\end{equation}
where $u$ is the (dereddened) apparent $u$-band magnitude of the object, $E(u)$ is the UV excess computed as in Eqs.\,\ref{eqn:uv_exc_ug}-\ref{eqn:uv_exc_ur} and $F_u^0$ is the SDSS $u$-band zero-point flux ($F_u^0$ = 3767.2~Jy).

We used Eq.\,\ref{eqn:fu_exc} \citep[cf. also][]{rigliaco2011} to measure the $u$-band flux excess over the whole sample and then converted $F_u^{exc}$ to $L_u^{exc}$ using a distance of 760 pc. 

\begin{figure}
\resizebox{\hsize}{!}{\includegraphics{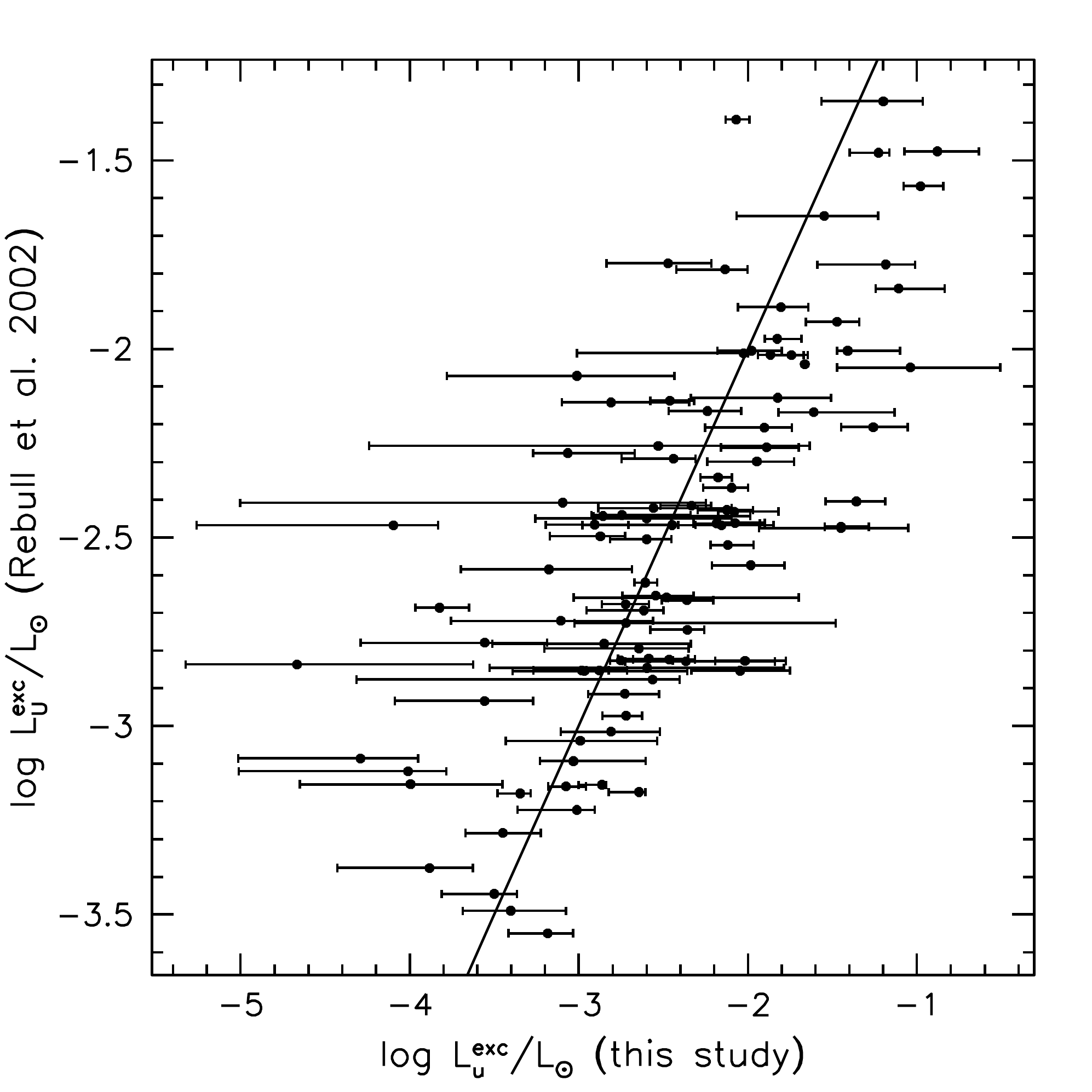}}
\caption{UV excess luminosity measurements obtained in this study are compared to \citet{rebull2002}'s for objects in common to both samples (L. Rebull, private communication). On the x-axis, dots correspond to the median L$_u^{exc}$ detected during the 2 week-long CFHT monitoring, while variability bars depict the actual range in L$_u^{exc}$ values measured at different observing epochs during the survey. The equality line is traced as a solid line to guide the eye.}
\label{fig:Luv_comparison}
\end{figure}

The results of a similar, single epoch photometric survey to detect and measure mass accretion in NGC~2264 from the UV excess diagnostics have been reported in \citet{rebull2002}. The authors measured the UV excess displayed by accreting stars from U-band and V-band photometry, by comparing the observed colors with colors expected based on the spectral type of individual objects. In Fig.\,\ref{fig:Luv_comparison}, we compare their measured UV excess luminosities (L. Rebull, private communication) with both the median L$_u^{exc}$ values and the relevant variability ranges we detect here, on a timescale of a couple of weeks, for CTTS targeted in both surveys (about 100). As can be observed, data points result to be well distributed around the equality line; the rms scatter measured about the line, amounting to 0.5~dex, is quite consistent with the average amount of variability on L$_u^{exc}$ (about 0.6~dex) detected across the sample from CFHT monitoring.

\subsection{Accretion luminosity and mass accretion rates} \label{sec:macc}
The $u$-band excess luminosity represents a single component of the total excess emission over the whole spectral range. In order to characterize the properties of the accretion process, it is therefore necessary to also take into account the flux excess at unobserved wavelengths, i.e., to apply a bolometric correction.

\citet{gullbring1998} showed that a direct correlation exists between the total accretion luminosity L$_{acc}$ and the excess luminosity observed in the U-band. This is indeed expected, since the excess emission likely dominates the observed flux at wavelengths $<$~4500~\AA \mbox{ }(i.e., below the $g$-band filter). In order to achieve this result, they measured the excess emission from spectra covering the range \mbox{3200--5400 \AA }, for a sample of 26 CTTS. They modeled the excess spectrum from the accretion spot assuming a region of constant temperature and density (slab model) and used this model fit to estimate the amount of excess emission at unobserved wavelengths. The authors eventually resolved to adopt a uniform correction factor, across the whole sample, for the ratio of total-to-observed excess flux, corresponding to a T$\sim$10\,000\,K. For each object, they subsequently synthetized a U-band magnitude from the observed spectrum and computed the corresponding flux excess. The comparison of the two quantities (L$_{acc}$ and L$_U^{exc}$) enabled the authors to yield a well-determined calibration equation, $\log(L_{acc}/L_\odot)=1.09_{-0.18}^{+0.04}\,\log(L_U^{exc}/L_\odot)+0.98_{-0.07}^{+0.02}$, that was ultimately consistent with the model predictions of \citet{calvet1998}, who introduced a detailed description of the accretion shock region.

Here we adopt a simple phenomenological approach to derive a direct relationship between the $u$-band excess luminosity and the total accretion luminosity L$_{acc}$. In a first approximation, we describe the stellar emission as a blackbody emission at the photospheric temperature T$_{phot}$. Similarly, the emission from the accretion spot is approximated by a blackbody at T=T$_{spot}$. The excess emission linked with accretion is thus given by the difference between the flux emitted by the distribution of accretion spots and the flux emitted by a corresponding area, of the same extension, on the stellar photosphere. In terms of blackbodies, this becomes the difference between a blackbody flux density at T=T$_{spot}$ and a blackbody flux density at T=T$_{phot}$, both integrated over the same area. In this picture, the ratio r of total-to-observed flux excess in the $u$-band, i.e. the corrective factor to be introduced to derive the total L$_{acc}$ from L$_u^{exc}$, can be estimated as 
\begin{equation}\label{eqn:u_exc_BB}
r=\frac{\sigma T_{spot}^4 - \sigma T_{phot}^4}{\pi\displaystyle\int_u{(B_{T_{spot}}(\lambda)-B_{T_{phot}}(\lambda))\,d\lambda}}\,,
\end{equation}
where the integration is performed over the $u$-band window. No assumptions on the filling factor are needed in this description.

In order to investigate the relationship between L$_u^{exc}$ and L$_{acc}$, we selected a subsample of 44 CTTS, covering the spectral type range M3.5:K2 (or equivalently the mass range $\sim$0.2--1.8 M$_\odot$) and whose variability, monitored in the $u$-band ($\lambda_{eff}$=3557\,\AA) and $r$-band ($\lambda_{eff}$=6261\,\AA) at CFHT, is likely accretion-dominated. An extensive spot modeling \citep[cf.][]{bouvier1993} of the simultaneous $u+r$ variability amplitudes has been performed in order to assess the dominating features of the variability displayed by individual objects; accretion-dominated sources have been identified as well-represented variables in terms of a surface spot distribution $\sim$$10^3$ to several~$\times10^3$\,K hotter than the stellar photosphere.  A full description of the variability analysis for the CFHT NGC~2264 members will be provided in a forthcoming paper (Venuti et al., in prep.). 

For each of the selected objects, we adopted the individual T$_{spot}$ estimate inferred from spot models, that corresponds, for a given object, to the color temperature of the spot distribution that best reproduces the simultaneous flux variations observed at different wavelengths; T$_{phot}$ is derived from the spectral type as described in Sect.~\ref{sec:parameters}. We then derived, for each object, the ratio r of Eq.\,\ref{eqn:u_exc_BB}, through a numerical integration of B($\lambda$) in the spectral range \mbox{3257--3857 \AA } \citep[cf.][]{F96SDSS}; an uncertainty on the value of r has been inferred correspondingly to the uncertainty on the spot temperature T$_{spot}$. Based on these estimates, the percentage of the total excess flux detectable in the $u$-band is on average $\sim$10\%. The accretion luminosity L$_{acc}$ is then computed as $L_{acc}=r\,L_u^{exc}$. 

A least-squares fit to the $\log\left(L_u^{exc}/L_\odot\right)$ vs. $\log\left(L_{acc}/L_\odot\right)$ distribution inferred as previously described provided us with the following calibration relationship:
\begin{equation}\label{eqn:Lacc_Lu}
\log\left(\frac{L_{acc}}{L_\odot}\right)=(0.97\pm0.03)\,\log\left(\frac{L_u^{exc}}{L_\odot}\right)+(1.09\pm0.07)\,
\end{equation}

\begin{figure}[b]
\centering
\includegraphics[width=8cm]{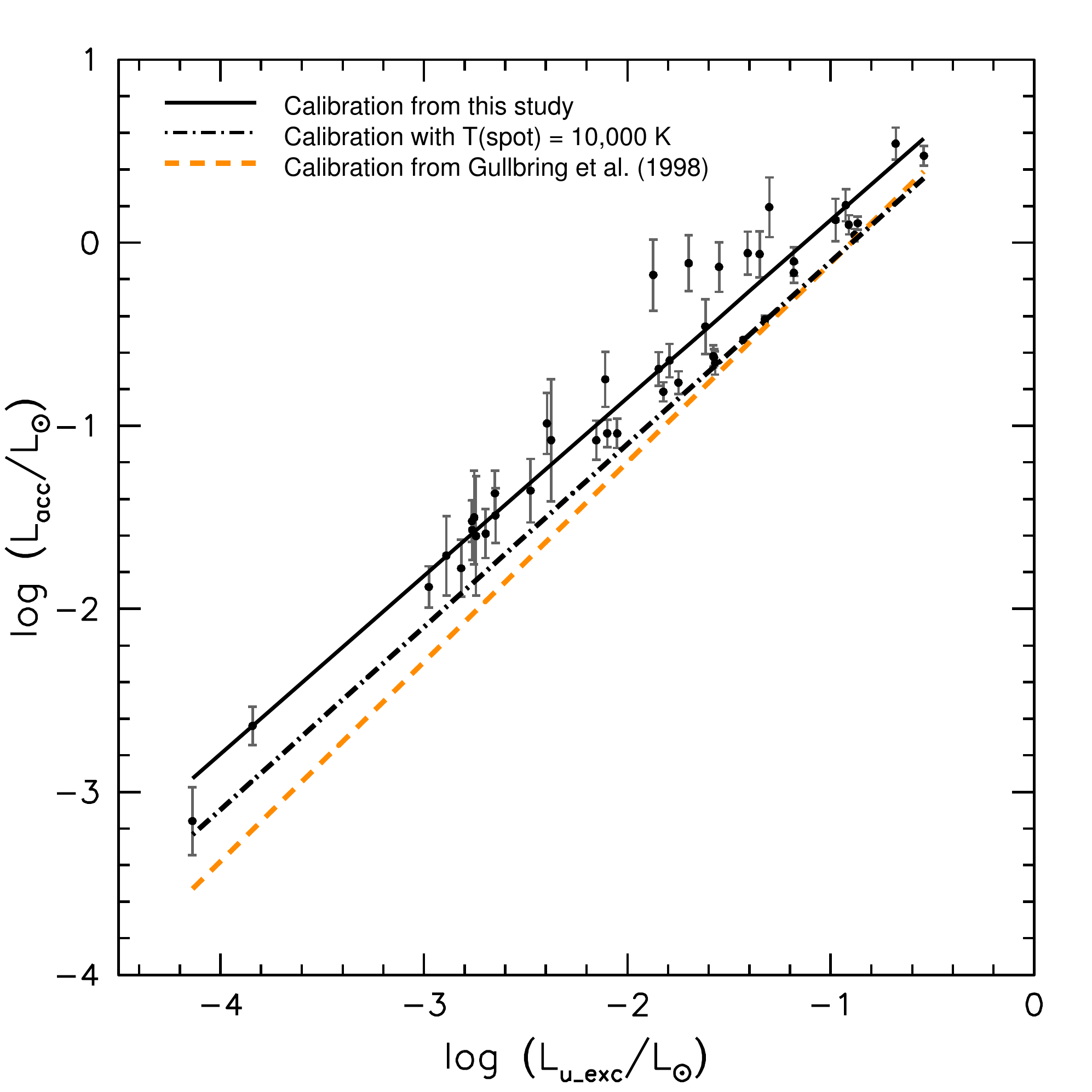}
\caption{L$_{acc}$--L$_u^{exc}$ calibration relationship. The black solid line traces the relationship derived in this study, while the orange dashed line represents the relationship by \citet{gullbring1998}. The data points represent the $u$-band excess luminosity and the accretion luminosity estimates inferred on an individual basis for a sample of 44 CTTS in our survey (see text). The error bars reflect the uncertainty on the individual T$_{spot}$ estimates. The black dashed-dotted line shows the calibration line that would be inferred, with the procedure followed in this study, in the assumption T$_{spot}$=10\,000\,K.}
\label{fig:cal_line_lacc}
\end{figure}

Fig.\,\ref{fig:cal_line_lacc} compares our result with the calibration inferred from the study of \citet{gullbring1998}; the parameters of the two calibrations are fairly consistent within the error bars. 
The locus traced by our point distribution is globally consistent with the prediction of \citeauthor{gullbring1998}'s study, albeit showing a systematic shift above their calibration line. Our estimates result to be larger by a factor of $\lesssim$2 to 4, corresponding to a difference from  $\sim$0.2~dex at the highest $\dot{M}_{acc}$ regimes ($\sim$$10^{-7}$ M$_\odot$/yr) to $\sim$0.6~dex at the lowest $\dot{M}_{acc}$ regimes ($\sim$$10^{-10}$ M$_\odot$/yr).

This difference between the two conversions is significantly reduced if we adopt, in our approach, a constant spot temperature of 10\,000\,K, an assumption closer to the uniform correction, at T$\sim$10\,000\,K, that was introduced in \citeauthor{gullbring1998}'s method. Indeed, individual T$_{spot}$ estimates adopted in our study cover a wide range of values, with a mean around $\sim$6500\,K, a spread of $\sim$2000--3000\,K at a given stellar temperature and nominal T$_{spot}$ that can reach up to $\sim$9000--11\,000\,K and down to $\sim$5000\,K. As described earlier, spot temperatures, in our approach, are inferred case by case by fitting simultaneous optical ($r$-band) and UV ($u$-band) variability amplitudes with spot models. Hence, these correspond to the color temperatures that best reproduce the observed, $\lambda$-dependent contrast of the spot distribution with the stellar reference emission level (which is the effect we aim at accounting for when introducing a bolometric correction). As the peak of the emission of a blackbody in the $u$-band occurs around 10\,000\,K, lower T$_{spot}$ estimates naturally imply that a larger correction factor is needed in order to recover the total flux emitted by the spot. While, for a typical photosphere (T$\sim$4000\,K), 12\% of the excess emission of a 10\,000\,K spot is observed in the $u$-band, this percentage reduces to $\sim$10\% for a 7000\,K spot and to less than 8\% for a 6000\,K spot. In order to evaluate how much of the discrepancy could actually be due to this effect, we repeated our procedure assuming a constant T$_{spot}$\,=\,10\,000\,K over the whole subsample of objects; as shown in Fig.\,\ref{fig:cal_line_lacc}, this indeed would allow us to mostly cancel the offset between the two calibrations. 

We are aware of the simplified nature of the blackbody approximation for stellar and spot fluxes; at the same time, we stress that the results inferred from this description are not inconsistent with those obtained from more sophisticated pictures. The discrepancy, albeit systematic, between L$_{acc}$ estimates from Eq.\,\ref{eqn:Lacc_Lu} and those from \citet{gullbring1998}'s calibration is smaller than, or comparable to, the usual estimates of the uncertainty affecting the accretion rate measurements ($\sim$0.5~dex), and a non-negligible contribution to this offset may actually derive from the different assumption on the correction factors adopted across the sample of objects considered in each study. Fig.\,10 of the theoretical study performed by \citet{calvet1998}, though it provides full endorsement of \citeauthor{gullbring1998}'s relationship, yet depicts the complexity of achieving a detailed physical representation of the accretion shock. The range of predictions from model configurations they explored would lie, on Fig.\,\ref{fig:cal_line_lacc} of our work, across the region delimited by our (Eq.\,\ref{eqn:Lacc_Lu}) and \citet{gullbring1998}'s relationship.

Considering this, and the fact that each model has its own limitations, we believe it beneficial to explore whether, and to which extent, information deduced from data might actually be sensitive to different model assumptions. We therefore adopt the calibration in Eq.\,\ref{eqn:Lacc_Lu} to estimate the total L$_{acc}$ from the measured $u$-band luminosity excess, and will refer at the same time to \citeauthor{gullbring1998}'s, where relevant, for comparison purposes.

If we assume that the accretion energy is reprocessed entirely into the accretion continuum \citep[cf.][]{herczeg2008}, the mass accretion rate $\dot{M}_{acc}$ can be derived from L$_{acc}$ as
\begin{equation}\label{eqn:Macc}
\dot{M}_{acc}=\left(1-\frac{R_*}{R_{in}}\right)^{-1}\frac{L_{acc}R_*}{GM_*}\sim 1.25 \, \frac{L_{acc}R_*}{GM_*},
\end{equation}
where M$_*$ and R$_*$ are the parameters of the central star, R$_{in}$ is the inner disk radius and we are assuming that the accretion funnel originates at the truncation radius of the disk, at R$_{in}\sim$ 5\,R$_*$ \citep{gullbring1998}. 

\begin{figure}[h]
\centering
\includegraphics[width=8.5cm]{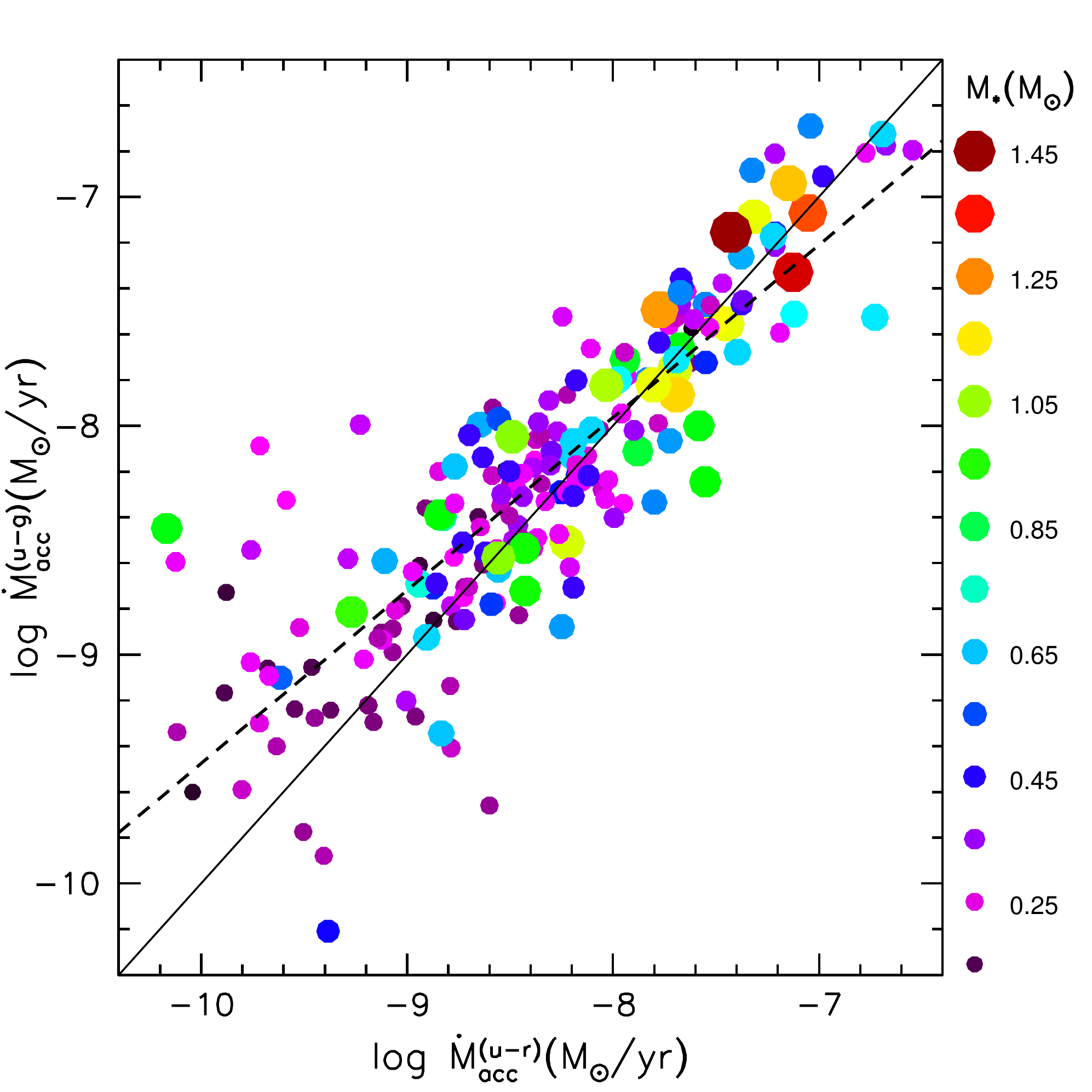}
\caption{$\dot{M}_{acc}$ values obtained with the two methods described in the text. $\dot{M}_{acc}$ values on the y-axis are inferred from $ugr$ snapshot photometry, while $\dot{M}_{acc}$ values on the x-axis correspond to the median UV excess detected during the $ur$ monitoring. Colors and sizes of symbols are scaled according to stellar mass M$_*$. The solid line marks the equality line; the linear regression to the point distribution is traced as a dashed line.}
\label{fig:macc_comp}
\end{figure} 

Fig.\,\ref{fig:macc_comp} compares the $\dot{M}_{acc}$ distributions inferred from each of the two UV derivations described at the beginning of Sect.\,\ref{sec:accretion}. The diagram reports the $\dot{M}_{acc}$ determinations for every accreting object, in the sample, that displays a UV excess according to the identification procedure previously described. The two distributions are on average consistent, within a small mass-dependent offset likely reflecting a residual overestimation of the $(u-g)$ color excess from Fig.\,\ref{fig:gr_ug_cand} (see discussion around Eq.\,\ref{eqn:uv_exc_ug}). Variability is likely to affect, at least partly, the scatter observed within $\pm$0.25~dex around the fitting line (see Sect.\,\ref{sec:macc_var}).

\begin{figure*}
\includegraphics[width=14cm]{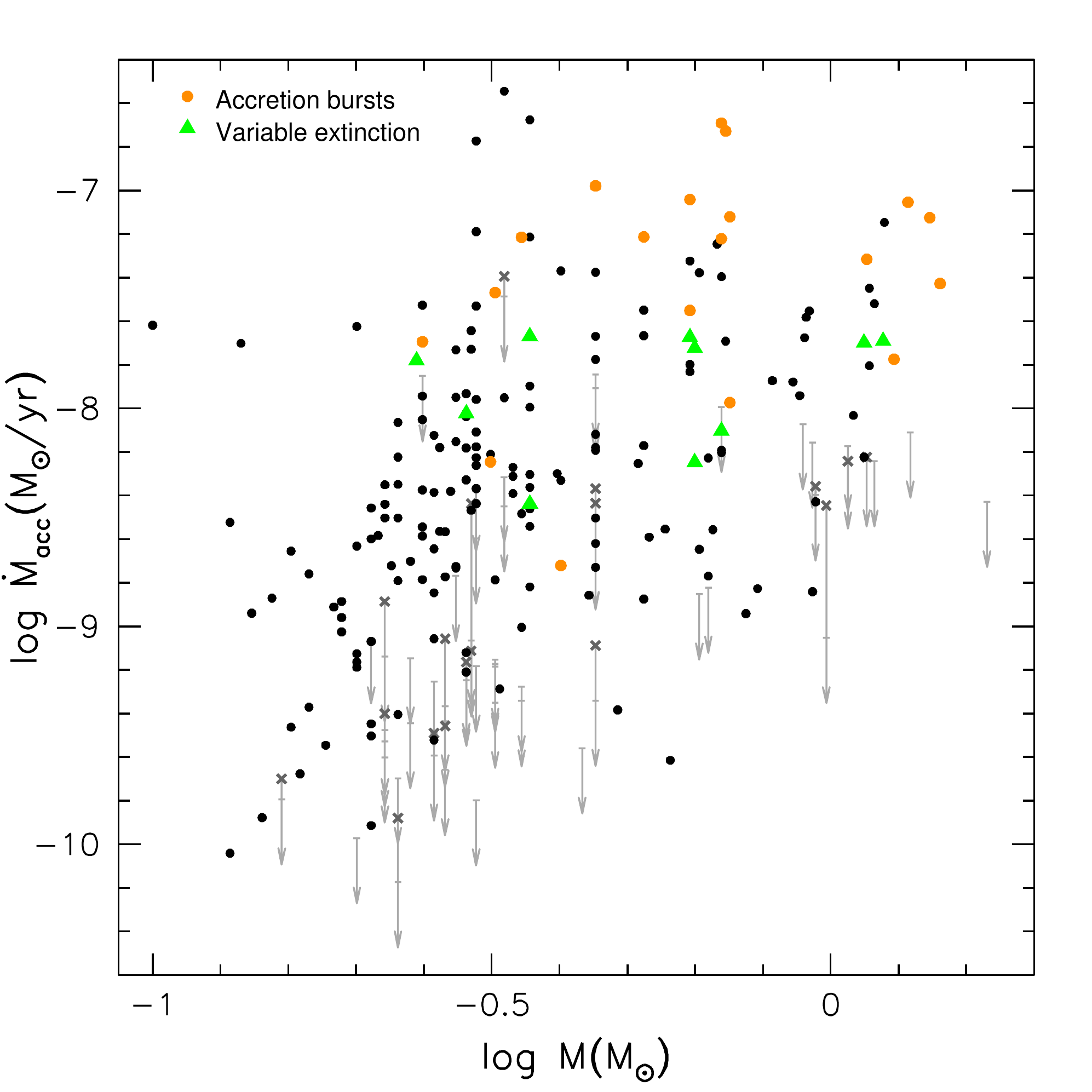}
\centering
\caption{$\dot{M}_{acc}$ distribution as a function of stellar mass for the population of NGC 2264 accreting members observed at CFHT. Upper limits are shown for those objects that fall below the estimated confidence threshold for accretion detection (see text). For some objects, no significant $\dot{M}_{acc}$ has been detected at the median state of the system, but they do show a significant accretion activity (i.e., above the detection level) at the brightest state; in these cases, an upper limit is placed at the detection threshold, while a cross marks the maximum $\dot{M}_{acc}$ actually detected. Orange dots and green triangles mark two subgroups of objects dominated respectively by stochastic accretion bursts and variable extinction from a rotating inner disk warp \citep{stauffer2014}.}
\label{fig:macc}
\end{figure*}

Fig.\,\ref{fig:macc} shows the mass accretion rate distribution inferred for a population of 236 high-confidence NGC~2264 members, as a function of stellar mass, whose range extends down to $\sim$0.1~M$_\odot$ and up to $\sim$1.5~M$_\odot$. These objects have been classified as CTTS, i.e., accreting objects, as opposed to WTTS among the PMS population analyzed so far, based on spectroscopic criteria (H$_\alpha$ EW, H$_\alpha$ width at 10\% intensity) and/or photometric criteria (significant UV excess from the CFHT sample, $\alpha_{IRAC}$ value consistent with Class II sources from the data of \citealp{teixeira2012}, large variability from the CFHT photometry; see Venuti et al., in prep., for this last point). The accretion parameters for these objects are reported in Table \ref{tab:accr_par}. The nominal values of the accretion rates are computed from the photometry corresponding to the median UV excess detected during the CFHT/MegaCam 2-week long monitoring and are estimated to be accurate within a factor of $\sim$3 typically. 

A confidence detection threshold for $\dot{M}_{acc}$ has been computed by evaluating the average scatter of the WTTS around the reference color sequence and testing the corresponding level of ``noise'' on the UV excess -- and consequently accretion rate -- determination. Namely, the magnitude-dependent rms scatter of WTTS around the reference sequence has been measured to compute a (magnitude-dependent) minimum UV excess level to be detected in order to be confident that accretion is indeed the dominant source of excess. This allowed us to trace a conservative boundary separating the objects with an unambiguous $\dot{M}_{acc}$ detection from the objects for which it is not possible, a priori, to evaluate the relevance of spurious contributions (i.e., chromospheric activity) to the apparent UV excess. For the accreting objects for which the actually measured UV excess is less than the corresponding boundary value, only an upper limit at the detection threshold has been reported for $\dot{M}_{acc}$.

\begin{table*}
\caption{UV excess, $u$-band excess luminosity, $\dot{M}_{acc}$ and $\dot{M}_{acc}$ variability estimates for NGC~2264 accreting members.}
\label{tab:accr_par}
\centering
\begin{tabular}{c c c c c c c c}
\hline\hline
  \multicolumn{1}{c}{Mon-ID} &
  \multicolumn{1}{c}{$(u-g)$ exc.} &
  \multicolumn{1}{c}{$(u-r)$ exc.\tablefootmark{1}} &
  \multicolumn{1}{c}{{$\log$ L$_u^{exc}$/L$_\odot$\tablefootmark{1}}} &
  \multicolumn{1}{c}{$\log\dot{M}^{(u-g)}_{acc}$} &
  \multicolumn{1}{c}{$\log\dot{M}^{(u-r)}_{acc}$\tablefootmark{1}} &
  \multicolumn{1}{c}{$\log\dot{M}_{max}/\dot{M}_{med}$\tablefootmark{2}} &
  \multicolumn{1}{c}{$\log\dot{M}_{med}/\dot{M}_{min}$\tablefootmark{3}} \\
\hline
  Mon-000007 & -0.16 & -0.65 & {-1.50} & -7.17 & -7.22 & 0.32 & \\
  Mon-000011 & -0.97 & -1.95 & {-0.88} & -7.53 & -6.73 & 0.24 & 0.19\\
  Mon-000017 & 0.03 & -0.27 & {-2.14} & & -8.24\tablefootmark{*} & 0.02 & \\
  Mon-000028 & -1.79 & -1.40 & {-3.38} & -8.90 & -9.13 & 0.40 & 0.59\\
  Mon-000040 & -0.46 & -0.34 & {-3.89} & -9.24 & -9.37 & 0.20 & 0.55\\
  Mon-000042 & -0.17 & -0.01 & {-5.08} & -9.34 & -9.48\tablefootmark{*} &  & \\
  Mon-000053 & -0.35 & -0.11 & {-4.31} & -9.06 & -9.46 & 0.33 & \\
  Mon-000056 & 0.04 & 0.01 &  &  & -8.24\tablefootmark{*} &  & \\
  Mon-000059 & -1.91 & -1.11 & {-3.32} & -8.36 & -8.91 & 0.26 & 0.57\\
  Mon-000063 & -0.53 & -1.03 & {-2.56} & -8.62 & -8.21 & 0.15 & \\
\hline
\end{tabular}
\tablefoot{A full version of the Table is available in electronic form at the CDS. A portion is shown here for guidance regarding its form and content.\\
\tablefoottext{1}{Value measured at the median luminosity state of the system.}\\
\tablefoottext{2,3}{Variability range detected around the median $\dot{M}_{acc}$ level during the 2 week-long $ur$ monitoring. If no accretion activity is detected above the detection limit for the object at the minimum/median luminosity level, but some significant accretion is detected at the brightest state, only the upper variability bar is reported (i.e., $\log\dot{M}_{max}/\dot{M}_{med(upper)}$); if the object is a non-detection in accretion, no variability measurement is reported.}\\
\tablefoottext{*}{Upper limit.}
}
\end{table*}

Little extensive investigation of mass accretion rates in NGC~2264 has been reported in the literature prior to the survey we are currently performing. Here we refer again to the study of \citet{rebull2002}. Those authors reported $\dot{M}_{acc}$ estimates for about 75 CTTS in the region; their accretion rates are derived from the measured UV excess luminosities by adopting \citet{gullbring1998}'s relationship to obtain the total L$_{acc}$ values. Results of the comparison of our $\dot{M}_{acc}$ estimates with theirs, for the few tens of sources in common (that cover nonetheless about 2 orders of magnitude in $\dot{M}_{acc}$), are quite consistent with what deduced from the comparison of the measured L$_{UV}^{exc}$ (Fig.\,\ref{fig:Luv_comparison}). When using the same L$_{u}^{exc}$--L$_{acc}$ calibration, $\dot{M}_{acc}$ estimates from the two studies distribute around the equality line, with an rms scatter of about 0.5~dex, pretty consistent with the average amount of $\dot{M}_{acc}$ variability we detect on week timescales during our monitoring (see Sect.\,\ref{sec:macc_var}).

Some important features can be observed on Fig.\,\ref{fig:macc}. The $\dot{M}_{acc}$ distribution globally traces a mass-dependent trend, showing an average increase of the mass accretion rates with stellar mass. The existence of such a trend has been investigated from observations over the last decade, albeit with a large uncertainty on the quantitative estimate of the actual relationship, which is likely to be dependent, among other factors, upon the probed stellar mass range and the mean age and age spread across the cluster (cf., e.g., \citealp{hartmann2006} and references therein, \citealp{rigliaco2011}, \citealp{manara2012}). Scattering of objects at a given stellar mass (see discussion later in the Section) also contributes to increasing the uncertainty on the actual relationship. Another important point to this respect is to understand the actual role played by detection biases in the observed mass--$\dot{M}_{acc}$ trend. As previously discussed, the statistical sample of accretion rates is limited at the lower $\dot{M}_{acc}$ end (Fig.\,\ref{fig:macc}), due to the impossibility of an accurate detection below a certain threshold; for this group of objects, only an upper limit to the actual $\dot{M}_{acc}$ value can be inferred. Contrast effects favor the detection of smaller accretion rates on cooler, i.e. lower-mass stars over higher-mass stars (as can be see in Fig.\,\ref{fig:macc}), thus unavoidably introducing a mass dependence. It is thus of interest to assess the degree of disentanglement between the observed mass--$\dot{M}_{acc}$ dependence and the effects of censoring in the data. 

A statistical approach that allows us to infer a robust measure of correlation in a given sample, in presence of multiple censoring (both ties, here points at the same x-value but with different y-values, and non-detections, i.e. upper limits), is a generalization of Kendall's $\tau$ test for correlation \citep[see][]{feigelson_babu, helsel}. The test consists in pairing the data points (measurements and upper limits) in all possible configurations and, for each pair, measuring the slope (positive or negative) of the line connecting the two points. A pair of data points having the same x-value will be counted as an indeterminate relationship, likewise a pair of upper limits; upper limits will uniquely contribute a definite relationship when paired with actual detections at higher y-value. The number of times positive slopes occur is then compared to the number of negative slopes, and their difference weighted against the total number of available pairs. This defines the statistical correlation coefficient, $\tau$. ln this picture, upper limits have the effect of lowering the likelihood of a correlation, if this is present in the data, as they introduce a number of indeterminate relationships among the tested pairs. Conversely, no correlation will be found in a pure sample of upper limits, even though these trace a well-defined trend, since the number of pairs with definite relationship will be zero. In the null hypothesis of no correlation, $\tau$ is expected to follow a normal distribution centered around zero; by measuring the deviation from zero of the actually found value, in terms of $\sigma$, it is thus possible to assess a confidence level for the presence of correlation in a given sample. 

The application of this statistical tool to our data allowed us to establish the presence of a correlation ($\tau\sim0.28$) that appears to hold across the whole mass range studied here with a confidence of more than 6\,$\sigma$. The robustness of this result was tested against a similar search for correlation in randomly generated distributions of $\dot{M}_{acc}$ in the log range [-10.5\,,\,-6.5] for the same stellar population with the same individual detection thresholds. The $\tau$ distribution inferred from 100 iterations is peaked around $\tau$=0.00 (no correlation) with a $\sigma$ of 0.04; this suggests that indeed an intrinsic correlation is observed in our data to a significance of $>$\,6\,$\sigma$. We then inferred a statistical estimate for the slope following the approach of Akritas-Theil-Sen nonparametric regression \citep[see][]{feigelson_babu}: 
\begin{enumerate}
\item we derived a first guess for the slope from a least-squares fit to the data distribution and explored a range of $\pm$0.5 around this first estimate with a step of 0.005; 
\item for each value $x$, we subtracted the $x$M$_*$ trend from the initial $\dot{M}_{acc}$ distribution and repeated the Kendall's $\tau$ computation procedure on the distribution of residues thus inferred. 
\end{enumerate}
The best fitting trend is the one that, subtracted to the point distribution, produces $\tau$=0, while an error bar can be inferred corresponding to the interval of values that produce $\tau\,<\,\mid\sigma\mid$. We applied this procedure to both the mass--$\dot{M}_{acc}$ distribution inferred from the $u-r$ excess distribution (see discussion around Eq.\,\ref{eqn:uv_exc_ur}) and the one inferred from the $u-g$ excess distribution (see discussion around Eq.\,\ref{eqn:uv_exc_ug}), obtaining respectively:
\begin{equation}\label{eqn:macc_mass_ur}
\dot{M}_{acc}\propto M_*^{1.5\pm0.2}
\end{equation}
\vspace{-7mm}
\begin{equation}\label{eqn:macc_mass_ug}
\dot{M}_{acc}\propto M_*^{1.3\pm0.2}
\end{equation}
The two trends are consistent within 1\,$\sigma$; the smaller exponent in Eq.\,\ref{eqn:macc_mass_ug} compared to Eq.\,\ref{eqn:macc_mass_ur} may reflect the small, mass-dependent offset between the $\dot{M}_{acc}$ distributions inferred from the two methods (cf. Fig.\,\ref{fig:macc_comp}). A slope of 1.7$\pm$0.2, instead of the estimate in Eq.\,\ref{eqn:macc_mass_ur}, would be inferred if we converted $u-r$ excesses to L$_{acc}$ values following \citet{gullbring1998}. The two estimates are still consistent within 1\,$\sigma$; this comparison illustrates that a (small) range in values around the true slope may be induced by external factors such as the L$_{u}^{exc}$--L$_{acc}$ conversion.

The analysis of correlation described above refers to the homogeneous sample of Fig.\,\ref{fig:macc} as a whole. While the relationship we infer seems to be robust across the whole sample, this likely provides a general, average information on the global trend of $\dot{M}_{acc}$ with mass, not a punctual description of the trend. We notice, for instance, that little mass dependence is observed in the upper envelope of the $\dot{M}_{acc}$ distribution above M$_*$$\sim$0.25\,M$_\odot$. We attempted a more detailed investigation of the observed $\dot{M}_{acc}$ vs. M$_*$ relationship by probing separately two different, similarly populated mass bins, M$_*$$\leq$0.3\,M$_\odot$ and M$_*$$>$0.3\,M$_\odot$. In both subgroups, we found some evidence of correlation, albeit with different values and significance levels: a correlation with a slope of 3.3$\pm$0.7 was inferred, to the 4\,$\sigma$ level, in the lower-mass group, while a correlation with a slope of 0.9$\pm$0.4 was obtained for the higher-mass group with a significance of 2.4\,$\sigma$. These results might suggest that the dependence of $\dot{M}_{acc}$ on M$_*$ is not uniform across the whole M$_*$-$\dot{M}_{acc}$ range, but shallower at higher masses and higher accretion rates; on the other hand, a closer investigation of the M$_*$--$\dot{M}_{acc}$ relationship from separate, small subsamples is necessarily more uncertain than the assessment of a general correlation trend observed across the whole sample.

Another remarkable feature on Fig.\,\ref{fig:macc} is the large dispersion in the $\dot{M}_{acc}$ values observed at each stellar mass, covering a range of $\sim$2 to 3 dex. Similar results have been inferred from previous statistical surveys of accretion in a given cluster (see, e.g., the recent studies of \citealp{rigliaco2011} on $\sigma$~Ori and of \citealp{manara2012} on the ONC). This implies that only the average behavior of the actual trend of $\dot{M}_{acc}$ with M$_*$ can be deduced, and suggests that different accretion regimes may co-exist within the same young stellar population. To this respect, we highlight in Fig.\,\ref{fig:macc} the accretion properties of two, morphologically well-distinct, classes of YSOs in NGC~2264, both actively accreting. The first corresponds to the newly identified \citep{stauffer2014} class of objects whose light curves are dominated by accretion bursts. The second class includes objects whose light curves show periodic or quasi-periodic flux dips likely resulting from a rotating inner disk warp partly occulting the stellar emission \citep{bouvier2007a, alencar2010}. As can be seen in Fig.\,\ref{fig:macc}, both classes of objects show a level of accretion on average larger than the mean level of accretion activity detected across the population of NGC~2264; however, the two classes are well distinguished, with the first showing a mean accretion rate about three times larger than the second.

\subsection{Mid-term $\dot{M}_{acc}$ variability} \label{sec:macc_var}
The scatter in Fig.\,\ref{fig:macc} could in principle partly arise from the huge photometric variability exhibited by T Tauri stars, both on short-term (hours) and mid-term (days) timescales \citep[e.g.,][]{menard1999}. Short-term variability can affect the $\dot{M}_{acc}$ estimates inferred for single objects, but its contribution is likely negligible for our purposes (i.e., for a statistical mapping of accretion throughout the population of the cluster). On the other hand, mid-term variability, comprising the geometric effects linked with stellar rotation and the intrinsic accretion variability (timescales relevant for hot spots), might effectively blur the $\dot{M}_{acc}$ distribution inferred from the mapping of the region, since each object is observed at an arbitrary phase. 

In order to assess the contribution of mid-term variability to the $\dot{M}_{acc}$ spread, we probed the variability of the detected $\dot{M}_{acc}$, over timescales relevant to stellar rotation, by measuring the UV excess and correspondingly computing the mass accretion rate from each observing epoch during the CFHT $u$-band and $r$-band monitoring. On average, 16-17 points distributed over the 2-week long survey have been considered to probe variability. This procedure allowed us to robustly associate a range of variability to the individual average $\dot{M}_{acc}$, inferred for each object as described in Sect. \ref{sec:macc}. 

\begin{figure}
\resizebox{\hsize}{!}{\includegraphics{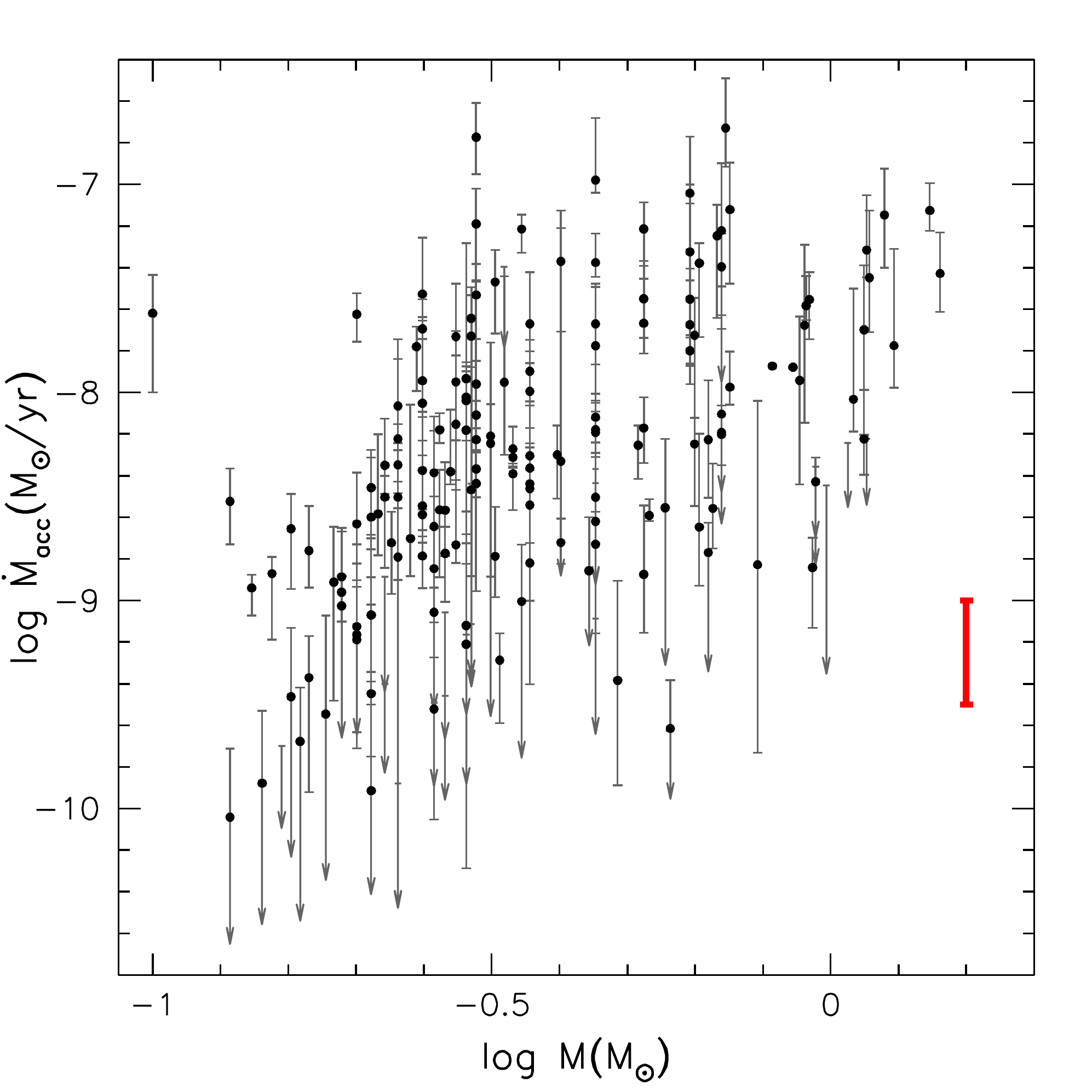}}
\caption{Variability of the mass accretion rates observed for NGC 2264 members over the 2-week long CFHT monitoring. Black dots correspond to the median $\dot{M}_{acc}$ measured for each object (see Fig.\,\ref{fig:macc}). For each object, the relevant variability bar is displayed, corresponding to the difference between the maximum and the minimum $\dot{M}_{acc}$ detected during the monitoring. A typical variability bar is shown in red.}
\label{fig:macc_var}
\end{figure}
Fig.\,\ref{fig:macc_var} shows the results inferred from this analysis; each variability bar encompasses all values of $\dot{M}_{acc}$ detected for the corresponding object during the monitoring (i.e., it traces the difference between the highest and the lowest detected $\dot{M}_{acc}$). Objects for which only an upper limit to the average $\dot{M}_{acc}$ had been derived have not been considered for this step of the analysis; for objects whose nominal $\dot{M}_{acc}$ value would fall below the detection threshold (Sect. \ref{sec:macc}) at certain epochs, an upper limit to the lowest apparent $\dot{M}_{acc}$ has been drawn. 

As can be observed on Fig.\,\ref{fig:macc_var}, variability properties are not uniform across the sample. No clear trends emerge with stellar mass or mass accretion rates; the geometry of individual systems might have a more important contribution in determining the apparent $\dot{M}_{acc}$ variability on week timescales, as suggested by the very few points on the diagram, maybe observed in a pole-on configuration, that show a negligible variation in the measured mass accretion rate. The (a)symmetry of individual bars, albeit potentially affected by the specific temporal sampling, may reflect the way accretion proceeds onto the central object, as well as the dominant features of the system: a strongly asymmetric bar with the median $\dot{M}_{acc}$ point close to the lower end can identify an object showing sudden, short-lived accretion bursts, while the opposite case well describes a system whose typical luminosity state is interspersed by short duration flux dips. These properties are indeed observed for some of the objects marked in Fig.\,\ref{fig:macc}. 

The detected amplitudes of variability in $\dot{M}_{acc}$ range from small fractions to $\sim$1 order of magnitude. In order to better visualize the average effects, we subdivided the sample of objects in Fig.\,\ref{fig:macc_var} in three broad mass intervals ($\mbox{M}_*<0.4\,\mbox{M}_\odot$; $0.4\,\mbox{M}_\odot \leq\mbox{M}_*\leq 1\, \mbox{M}_\odot; \mbox{ M}_*> 1\, \mbox{M}_\odot$) and for each of them computed the mean $\dot{M}_{acc}$ and amplitude of variability. These results are reported in Table \ref{tab:var_bars}, while a typical variability bar is represented in red in the lower right part of Fig.\,\ref{fig:macc_var}. On average, a variability of $\sim$~0.5 dex in $\dot{M}_{acc}$ is observed over a couple of weeks (which correspond to $\sim$2-3 rotational cycles for most of the objects; see \citealp{affer2013}), against a typical $\dot{M}_{acc}$ spread of $\sim$2 dex. This comparison clearly shows that the observed $\dot{M}_{acc}$ spread at each mass is significantly larger than what can be accounted for by variability on week timescales, be it intrinsic (non-steady accretion) or geometric (rotational modulation).
\begin{table}[b]
\caption{Average variability in $\dot{M}_{acc}$ displayed, on week timescales, by different, mass-sorted, groups of accreting members of NGC~2264.}
\label{tab:var_bars}
\centering
\begin{tabular}{l c c c}
\hline\hline
\multicolumn{1}{c}{{\bf M$_*$ group (M$_\odot$) :}} & {\bf $<$ 0.4} & {\bf 0.4--1} & {\bf $>$ 1 }\\
\hline
$<$M$_*$$>$ (M$_\odot$) & 0.26 & 0.60 & 1.20 \\
$<\log(\dot{M}^{med}_{acc})>$ (M$_\odot$/yr) & -8.36 & -7.92 & -7.58 \\
$<\Delta\log\dot{M}_{acc}\tablefootmark{1}>$ (dex) & 0.50 & 0.46 & 0.52 \\
$<\log{\frac{\mbox{bar}_{+}}{\mbox{bar}_{-}}}\tablefootmark{2}>$ & -0.1$\pm$0.3 & 0.0$\pm$0.3 & 0.1$\pm$0.2 \\
\hline
\end{tabular}
\tablefoot{Estimates only take into account objects for which a detection, and not an upper limit, is available for the lowest $\dot{M}_{acc}$ phase.\\
\tablefoottext{1}{$\log{\dot{M}^{max}_{acc}} - \log{\dot{M}^{min}_{acc}}$}\\
\tablefoottext{2}{Index probing the symmetry of the variability bars around $\dot{M}^{med}_{acc}$: bar$_{+}=\log{\dot{M}^{max}_{acc}} - \log{\dot{M}^{med}_{acc}}$; bar$_{-}=\log{\dot{M}^{med}_{acc}} - \log{\dot{M}^{min}_{acc}}$.}
}
\end{table}

\subsection{Accretion variability: intrinsic vs. modulated}
As discussed in Sect.\,\ref{sec:macc_var} (see also Sect.\,\ref{sec:spread_var}), an average variability of $\sim$0.5 dex is derived when monitoring the observed $\dot{M}_{acc}$ for individual objects over days-to-weeks timescales. This amount of variability is expected to incorporate both the intrinsic variability associated with the accretion process and the geometric effect of varying accretion shock projection along the line of sight as the star rotates.

\begin{figure}
\centering
\includegraphics[width=8.5cm]{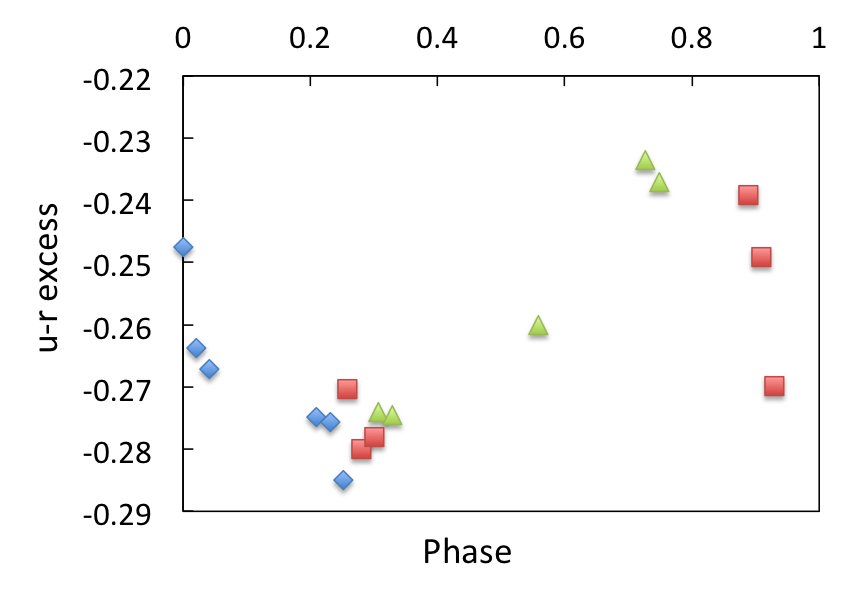}
\caption{Phased diagram of the variability of the UV excess monitored over a baseline varying from a few hours to a few rotational cycles for an NGC~2264 accreting member (M$_*$=1.13 M$_\odot$; $\dot{M}_{acc}$=5.98$\times10^{-9}$ M$_\odot$/yr; P$_{rot}$=4.767 days, from \citealp{affer2013}). The phase is computed starting from the epoch of the first measurement; the UV excess at each epoch is computed as described in Eq.\,\ref{eqn:uv_exc_ur}. Different symbols correspond to different rotational cycles.}
\label{fig:uv_exc_mod}
\end{figure}

An illustration of this is shown in Fig.\,\ref{fig:uv_exc_mod}.
The underlying sinusoidal profile describes the geometric modulation of the accretion spot linked with stellar rotation, which is reflected in the measured UV excess and, consequently, in the value of $\dot{M}_{acc}$ inferred at a given phase. Two additional effects contribute to the variability picture: the intrinsic variability of accretion on days timescales, reflected in the variation of the UV excess measured at the same phase in two different cycles, and the intrinsic variability of accretion on hours timescales, reflected in the rms variation of the UV excess measured within the same rotational cycle around a given phase (i.e., measurements obtained a few hours apart during the same night). 

In order to probe the relative significance of these different contributions to the monitored $\dot{M}_{acc}$ variability and characterize the intrinsic accretion variability over hours-to-days timescales, we selected a subsample of 35 accreting members of NGC~2264, subject to the following conditions: 
\begin{enumerate}
\item being an actual detection in $\dot{M}_{acc}$ (i.e., not an upper limit);
\item displaying a detectable periodicity, known from the study of \citet{affer2013};
\item having rotational period small enough to be able to detect at least a partial superposition of different cycles in the phased diagram.
\end{enumerate}
\citet{affer2013}'s study of rotation in NGC~2264 is based on a 23~day-long optical monitoring of the star-forming region performed with the CoRoT satellite in March 2008. Accreting stars with detected rotational modulation amount to about 16\% of the CTTS in our sample. These are distributed over the whole $\dot{M}_{acc}$ range at all masses in Fig.\,\ref{fig:macc}, except at the lowest $\dot{M}_{acc}$ regimes ($\lesssim 5\times10^{-10}$ M$_\odot$/yr), typically populated by objects fainter than 17 in $r$ and hence not attainable with CoRoT.

For each of these objects selected as described above, we phased the light curve of the UV excess and measured independently the average rms variation of detections obtained within the same observing night and the average rms variation between the average UV excess measured in a given phase bin of 0.05 at different rotational cycles.

The average peak-to-peak variation in the color excess throughout the sample amounts to $\sim$0.47 mag, which corresponds to a variation of $\sim$0.52 dex in $\dot{M}_{acc}$, consistent with what we found in Sect.\,\ref{sec:macc_var}. 50\% of the objects show an average intrinsic rms variation of $<$0.1 mag at a given phase from cycle to cycle (80\% at $\leq$0.15 mag and 90\% at $\leq$0.25 mag). Hence, on average a ratio of $\sim$0.25 between the intrinsic rms variation on day~timescales and the peak-to-peak variation can be inferred. The average rms variation on hour timescales is smaller, amounting to $\leq$0.05 mag. On the assumption that the two contributions (accretion variability on hours vs. days timescales) are mutually independent, we conclude that these two sources of variability together contribute on average just $\sim$28\% of the observed peak-to-peak variation in the UV excess. For a typical $\Delta\dot{M}_{acc}$ of 0.52 dex, intrinsic accretion variability is thus expected to contribute for about 0.15 dex, while a major contribution arises from the geometric modulation of the accretion shock. 

Interestingly, in the picture of modulation-dominated variability and assuming a homogeneous phase sampling, we would expect to statistically observe symmetric variability bars around the average value of $\dot{M}_{acc}$; this is indeed consistent with the results reported in Sect.\,\ref{sec:macc_var} and synthetized in Table \ref{tab:var_bars}.

\section{Discussion} \label{sec:discussion}
Accretion is a key ingredient in early stellar evolution: it regulates the star-disk interaction over the first few million years, with a major impact on the dynamical evolution of both the central object and the protoplanetary disk. Inferring a complete picture of accretion within large, well-characterized samples of young stars allows us to explore the accretion properties over a variety of parameters (e.g., stellar mass, age, disk properties), and hence represents a crucial issue for understanding how accretion evolves on an individual vs. statistical basis.

\subsection{The $M_*-\dot{M}_{acc}$ relationship}
Understanding what relationship links disk accretion rates and the mass of central objects has been a major focus for studies of accretion in YSOs over the last years. A prime motivation for this lies in that the M$_*$--$\dot{M}_{acc}$ relationship likely reflects the actual mechanisms governing the accretion process. 

\citet{hartmann2006} reviewed a number of different mechanisms which may be relevant to disk accretion and its variability and to the M$_*$--$\dot{M}_{acc}$ relationship. As noted by those authors, several unrelated mechanisms may contribute to establishing this relationship, and those mechanisms may act differently in different mass regimes. A number of previous studies have attempted to better understand the mass dependence of the M$_*$--$\dot{M}_{acc}$ relationship \citep{vorobyov2008, rigliaco2011, fang2013}; these analyses suggest a diminution of the steepness of the $\dot{M}_{acc}$--M$_*$ dependence as M$_*$ increases from a few tenths or hundredths of solar mass to a few solar masses. This is consistent with what we observe here for NGC~2264; on the other hand, the overall trend observed in the subsolar-to-solar mass regimes appears to be significantly shallower than that observed in the intermediate (1--10\,M$_\odot$) mass regime \citep{ercolano2014}.

In this study, we examined the distribution of mass accretion rates over mass for a homogeneous population of 236 confirmed, accreting members in NGC~2264, covering the mass range $\sim$0.1--1.5~M$_\odot$. We inferred the presence of a correlation between $\dot{M}_{acc}$ and M$_*$ and further showed that this result is not significantly affected by a mass-dependence in the $\dot{M}_{acc}$ detection limits. The same fraction of accreting stars over the total young population ($\sim$40\%) is roughly observed at lower ($<$0.4~M$_\odot$) and higher ($>$1~M$_\odot$) masses; unless the true fraction of accreting objects is considerably larger at higher masses than at lower masses, this attests that no significant selection biases towards lower-mass objects affect our accretion survey. Hence, we obtain $\dot{M}_{acc}$\,$\propto$\,M$_*^\alpha$, with $\alpha$$\sim$1.4$\pm$0.3 (Eqs.\,\,\ref{eqn:macc_mass_ur}-\ref{eqn:macc_mass_ug}). This exponent is smaller than the $\alpha$=2 dependence inferred by \citet{muzerolle2003} over the mass range 0.04--1\,M$_\odot$; the trend they obtained, however, is strongly influenced by the substellar component of the PMS population, which is not probed in our study. Indeed, results similar to \citeauthor{muzerolle2003}'s findings have been subsequently inferred from studies probing a comparable mass range in the substellar down to planetary mass regimes (e.g., \citealp{mohanty2005} for a composite sample of members from different star-forming regions, \citealp{natta2006} in $\rho$ Ohiuchi, \citealp{herczeg2008} in Taurus, \citealp{alcala2014} in Lupus). On the other end, studies addressing more specifically the T Tauri mass regime have recently proposed somewhat lower estimates for the quantitative dependence of $\dot{M}_{acc}$ on M$_*$. \citet{rigliaco2011} probed a mass range, in $\sigma$~Ori, similar to the one investigated in \citeauthor{muzerolle2003}'s study (0.06--1 M$_\odot$) and inferred a value of $\alpha=1.6 \pm 0.4$. \citet{barentsen2011} studied a population of 158 accreting stars, covering the mass range 0.2--2\,M$_\odot$, in IC~1396, and inferred a M$_*$--$\dot{M}_{acc}$ relationship with $\alpha=1.1\pm0.2$. All these results are quite consistent with the estimate of slope here inferred. The range of $\alpha$ from these studies may in part reflect the inhomogeneous nature of the methodologies and stellar samples; however, taken at face value, the results suggest a possible dependence of $\alpha$ on mass, with the higher mass regime showing a shallower slope.

\subsection{Accretion variability: what contribution to the $\dot{M}_{acc}$ spread at a given mass?}\label{sec:spread_var}
A ubiquitous result of the studies of $\dot{M}_{acc}$ distributions for large samples of young stars is the significant spread observed in $\dot{M}_{acc}$ at a given mass, enveloping the average trend over 2 to 3 orders of magnitude. CTTS are known to display a strong variability over different timescales; this takes contribution both from geometric effects linked with the rotational modulation of surface starspots and disk features \citep{herbst1994} and from the intrinsic variability of the accretion process. Variability could thus be partly responsible for the observed scatter. 

In order to probe this effect, we monitored the UV excess and $\dot{M}_{acc}$ variability for about 150 NGC~2264 accreting members over two full weeks, with typically 16-17 epochs for each object; this allowed us to show that variability on week timescales determines on average a variation of the detected $\dot{M}_{acc}$ over $\sim$0.5 dex, significantly smaller than the extent of the average $\dot{M}_{acc}$ scatter. 

This study complements the results inferred from recent studies that explored the variability in mass accretion rates on longer timescales. \citet{nguyen2009} analyzed the $\dot{M}_{acc}$ variability of a sample of 40 members in Taurus-Auriga and Chamaeleon I, from multi-epoch optical spectra distributed over $\sim$10 months, with typically 4 epochs per object on a baseline varying from hours to months. They probed the accretion properties from H$\alpha$ 10\% width and Ca II flux and inferred a median variability estimate amounting respectively to 0.65 dex and 0.35 dex. \citet{fang2013} probed the $\dot{M}_{acc}$ variability of several tens of members in L1641, monitored on timescales of 1, 10 and 22 months; measuring the amount of accretion from the full width of the H$_\alpha$ line at 10\% intensity, the authors inferred a typical $\dot{M}_{acc}$ variability of $\sim$0.6 dex.

A recent spectroscopic survey of the $\dot{M}_{acc}$ variability in T~Tauri and Herbig~Ae stars, addressing shorter ($\lesssim$hour to days) timescales, has been performed by \citet{costigan2014}. The authors probed accretion properties and their variability from H$_\alpha$~EW measurements for a sample of 14 objects; targets have been monitored over $\lesssim$ 1~hour blocks, in some cases iterated within a single night, and repeated over a few nights. Combining the results of this short-term variability monitoring with multi-year information for a few objects, as well as with results from different surveys, the authors found that the dominant contribution to the observed variability in $\dot{M}_{acc}$ arises from the days timescale, with a spread in accretion rates ranging from 0.04 to 0.4~dex.

All these results robustly show that variability on day-to-month timescales does not impact significantly on the large spread observed in $\dot{M}_{acc}$ around the average $\dot{M}_{acc}$--M$_*$ trend; hence, the spread is real and possibly linked to the different evolutionary stages of individual objects (see Sect.\,\ref{sec:discussion_spread}). Additionally, a strong overlap in the age distributions for the populations of accreting (CTTS) and non-accreting (WTTS) members seems to hold; modulo the issue of an accurate age determination for young stars, this suggests that an intrinsically large range in $\dot{M}_{acc}$ exists at a given mass and age, which may reflect various other parameters (like, e.g., the properties of the stellar magnetic field or the initial mass of the disk). 

Remarkably, the order-of-magnitude estimates for the $\dot{M}_{acc}$ variability inferred on day-to-week timescales (this study; \citealp{costigan2014}) and on month-to-year timescales \citep{nguyen2009, fang2013} are quite consistent; this suggests that the same mechanisms dominate variability over these baselines, with a major contribution arising from the mid-term variability component. This picture is supported by the results of the comparison, shown in Fig.\,\ref{fig:Luv_comparison}, between the L$_{UV}^{exc}$ measurements obtained from ours and \citet{rebull2002}'s survey of NGC~2264, respectively; what we observe in this case is that, indeed, UV excess luminosities measured at the two epochs are globally consistent, with an rms scatter, about the equality line, of the same order of magnitude as the average amount of variability detected during our 2 week-long CFHT monitoring.

\subsection{What is the origin of the intra-cluster spread in $\dot{M}_{acc}$?} \label{sec:discussion_spread}
A star-to-star variety in the evolutionary stage and evolutionary scenario may provide an important contribution to the interpretation of the large range of accretion properties observed within the region. The cluster shows a hierarchical structure (see \citealp{dahm08} and references therein), with regions of active, current star formation embedded in a more widespread population of somewhat older stars \citep{sung09}. Several episodes of star formation may have occurred, as proposed by \citet{adams1983}. Despite the uncertainty on a quantitative estimate, there is a general consensus about an age spread of up to several Myr within the population of NGC~2264. It is thus of interest to attempt a characterization of the accretion properties relative to age indicators and disk evolutionary stages. 

\begin{figure}
\centering
\includegraphics[width=8.7cm]{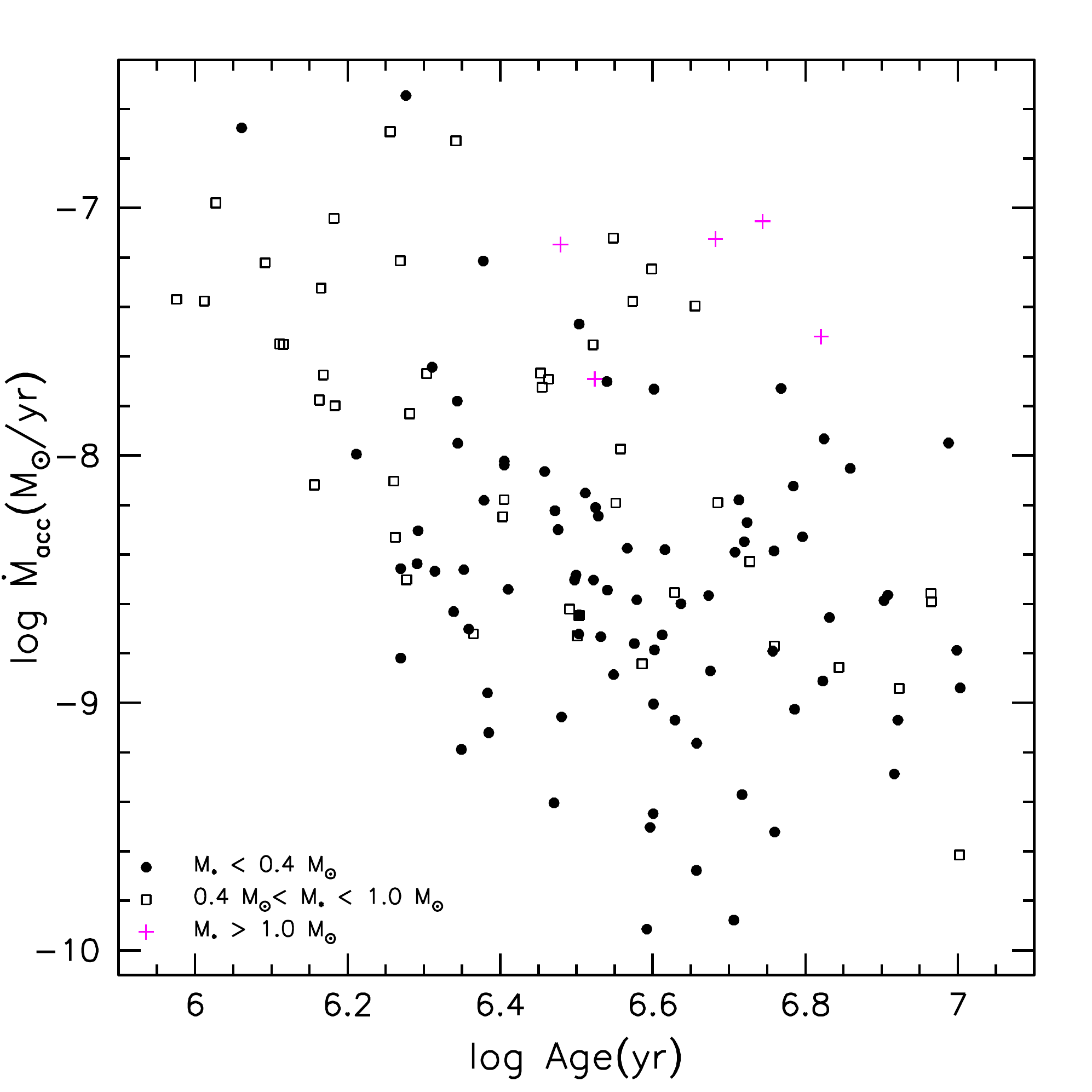}
\caption{Evolution of the accretion properties as a function of stellar age for NGC~2264 accreting members. Three different mass groups are compared: $M_*<0.4\,M_\odot$  (filled circles); $\,0.4\,M_\odot\leq M_* \leq 1\,M_\odot$ (open squares); $\,M_*>1\,M_\odot$ (crosses).}
\label{fig:macc_age}
\end{figure}

Looking at how points distribute on the HR diagram of the cluster (Fig.\,\ref{fig:HRD}) and comparing this to model isochrones provides a debated indication on nominal ages. In order to probe the evolution of accretion properties with stellar age, we inferred an age estimate for each source from isochrone fitting, adopting the models of \citet{siess2000}. The results for accreting sources are displayed vs. the $\dot{M}_{acc}$ estimates in Fig.\,\ref{fig:macc_age}. Only objects which are actual detections in $\dot{M}_{acc}$ and whose age estimate from models is between $\sim10^6$\,yr and $10^7$\,yr have been retained for the comparison. 

An overall decreasing trend of the average $\dot{M}_{acc}$ with increasing stellar ages is traced in Fig.\,\ref{fig:macc_age}, consistently with previous results regarding the average evolution of accretion with stellar age \citep[e.g.][]{hartmann1998, sicilia_aguilar2010, manara2012}. The anticorrelation trend is well described in the M$_*$\,$<$\,1\,M$_\odot$ regime, reaching a significance of over 5\,$\sigma$ according to a Spearman's rank order test \citep{numerical_recipes}. A linear regression to this group of objects yields
\begin{equation}\label{eqn:macc_age}
\dot{M}_{acc}\propto t^{-\eta},\,\eta\simeq1.51,
\end{equation}
quite consistent with the $\eta=1.5$ estimate inferred by \citet{hartmann1998}. Little information can be achieved on the accretion properties of the higher-mass group (M$_*$$>$1\,M$_\odot$) from our UV excess analysis.

The actual $\dot{M}_{acc}$--age relationship may vary depending on the mass range probed, as suggested by the looser correlation that can be inferred from the least massive group of objects (M$_*$$<$0.4\,M$_\odot$) compared to the 0.4 -- 1\,M$_\odot$ group. This might reflect truly different evolutionary timescales, but also be affected by uncertainties on the age estimates inferred from models. Indeed, a clear-cut decreasing trend is observed on the lower-left envelope of the data point distribution in Fig.\,\ref{fig:macc_age}, with the youngest group being among the most luminous sources in the sample. While limits on the detection of small accretion rates in luminous stars might somewhat affect the observed trend, a more significant bias here is likely introduced by the model age estimates. We notice that lower-mass stars appear to be systematically older than higher-mass stars, based on model ages. If this is not a physical effect, but rather an artifact driven by systematics in the model, the actual anticorrelation trend that can be derived from Fig.\,\ref{fig:macc_age} is, at best, doubtful.  

The reliability of such time evolution analyses of accretion is subject to the assumption that the interpretation of the luminosity spread for a given PMS population on the HR diagram in terms of a true age spread is overall correct. However, this interpretation is subject to a severe controversy. Indeed, recent studies have questioned the accuracy of individual age estimates from the stellar luminosity. \citet{baraffe2009} claimed that the luminosity spread observed on the HR diagram of a given star-forming region does not correspond to an actual age spread, but merely reflects the individual episodic accretion history, whose dynamics may mimic an age spread of up to a few Myr for coeval stars. While no consensus has obviously been achieved on this issue, an intrinsic age spread of the order of Myr, if present, is likely to be reflected in different properties of the region, such as the spatial distribution of members. This point is discussed in Sect.\,\ref{sec:macc_mapping}.

\begin{figure}
\centering
\includegraphics[width=8.7cm]{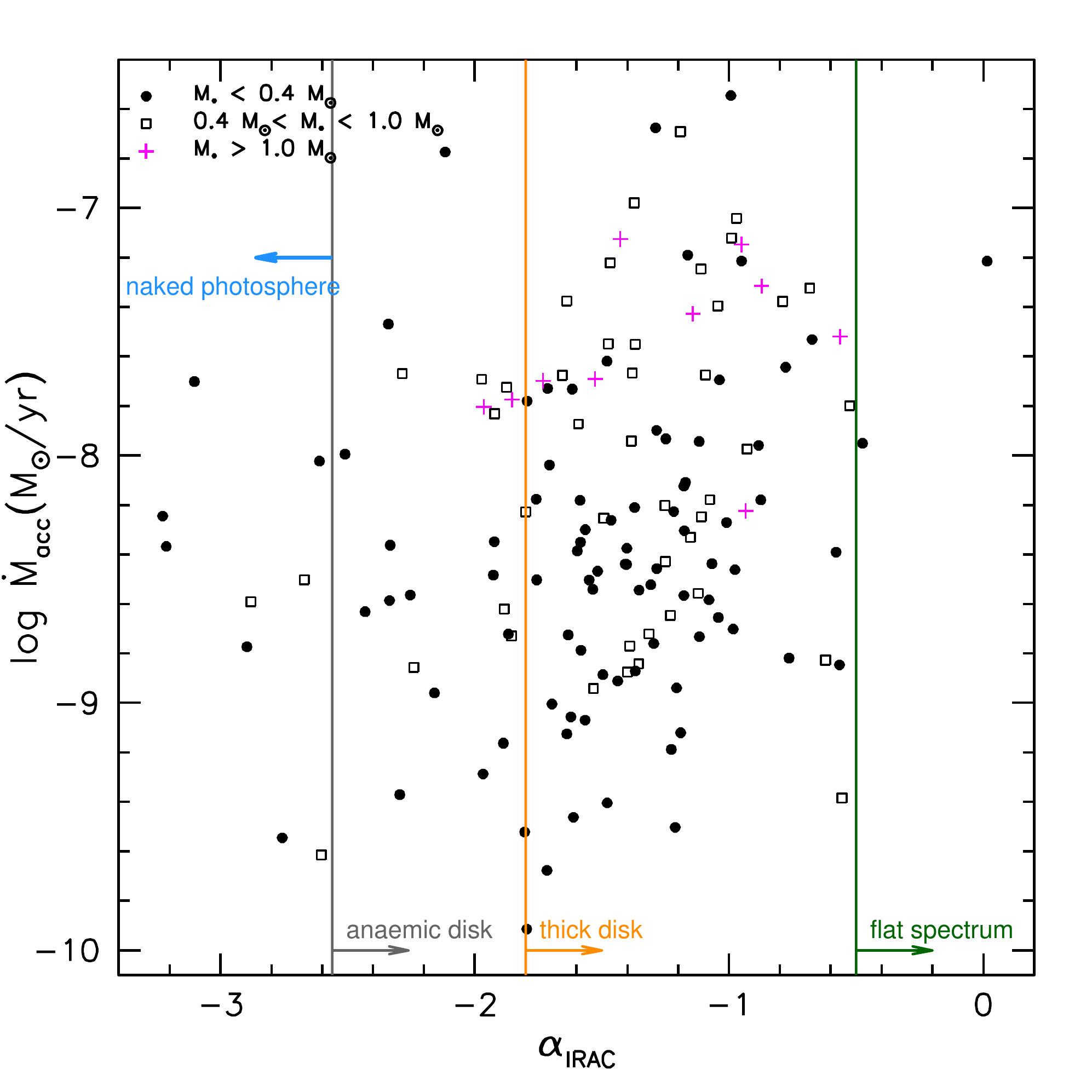}
\caption{$\dot{M}_{acc}$--$\alpha_{IRAC}$ relationship. The $\alpha_{IRAC}$-based disk morphology classification follows Table 2 of \citet{teixeira2012}. Three different mass groups are compared: $M_*<0.4\,M_\odot$  (filled circles); $\,0.4\,M_\odot\leq M_* \leq 1\,M_\odot$ (open squares); $\,M_*>1\,M_\odot$ (crosses).}
\label{fig:macc_alpha_irac}
\end{figure}

As mentioned in Sect.\,\ref{sec:acc_diag}, information on the circumstellar disk morphology, hence on the evolutionary status, is conveyed by the mid-IR slope of the SED of the system. In Fig.\,\ref{fig:macc_alpha_irac}, the accretion properties for three different mass groups of members are compared to the $\alpha_{IRAC}$ classification from \citet{teixeira2012}. Smaller values of $\alpha_{IRAC}$ correspond to more evolved disks; the average accretion rates decrease as disks become more evolved, albeit within a significant spread for each mass group. 

These comparisons show that several parameters, beyond stellar mass, regulate the accretion evolution of individual systems and eventually contribute to shape the complex picture of a star-forming region such as we may infer from an extensive accretion survey.

\subsection{A spatial mapping of accretion properties in NGC~2264} \label{sec:macc_mapping}
Characterizing the spatial distribution of accreting members according to their accretion properties is a useful tool to investigate the evolution dynamics of the region. 

\begin{figure}
\resizebox{\hsize}{!}{\includegraphics{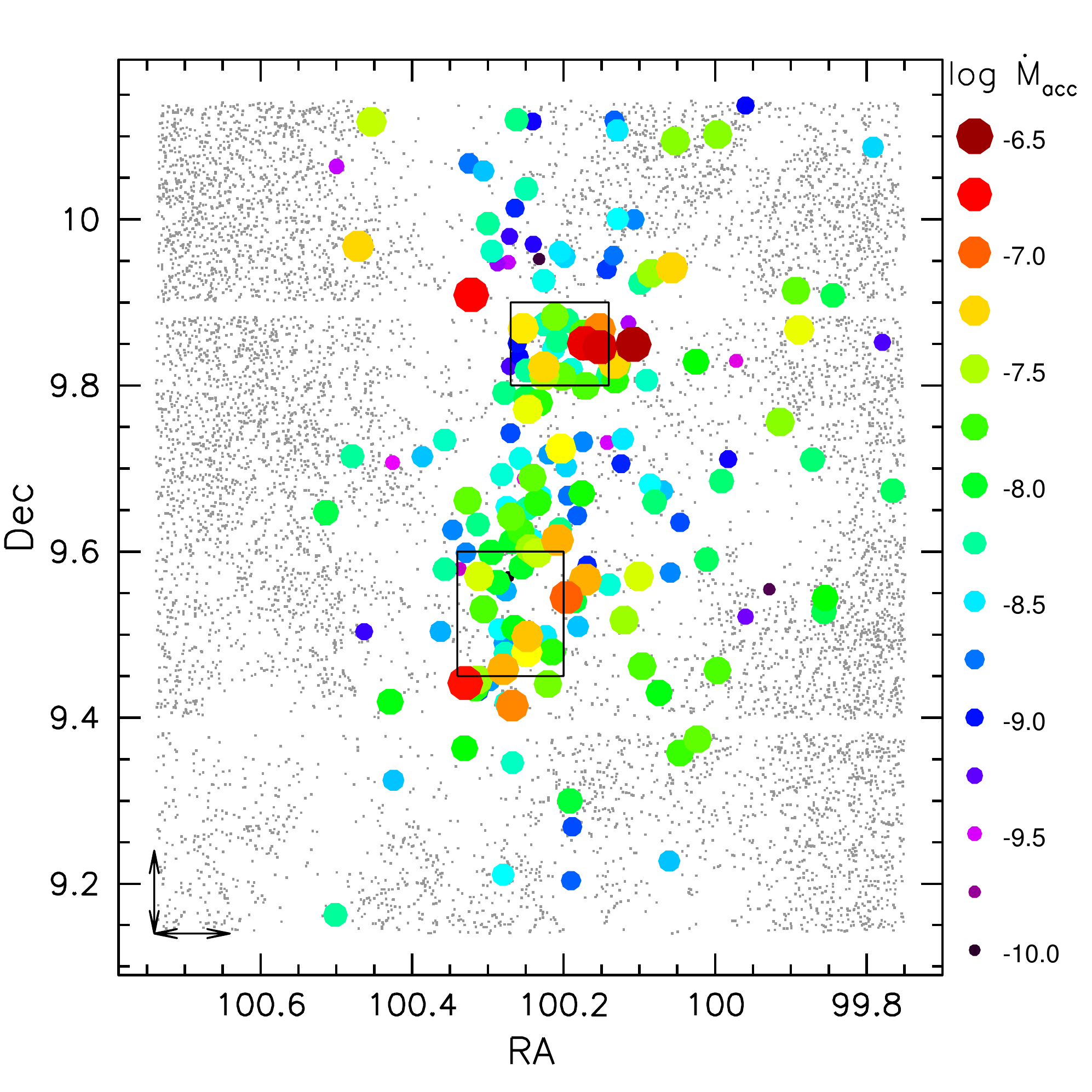}}
\caption{Spatial mapping of accretion properties in NGC~2264. Small grey dots indicate the distribution of field stars, large dots indicate accreting cluster members. Only objects with an actual $\dot{M}_{acc}$ detection are shown. Symbol colors and sizes are scaled according to the value of $\dot{M}_{acc}$. The black boxes mark the regions of maximum stellar density in the cluster \citep[from][]{lamm04}.  Double arrows mark a physical distance of 1.5 parsecs.}
\label{fig:macc_radec}
\end{figure}

Fig.\,\ref{fig:macc_radec} shows a spatial mapping of accretion in NGC~2264. Most accretors are projected onto the cluster; a part of the population of accreting or mildly accreting members are instead located on the peripheral regions. Some interesting features can be observed: the strongest accretors tend to be concentrated around the two known active sites of star formation within the cloud, a few parsecs apart, while the population of more moderate accretors tends to be more evenly distributed throughout the whole cluster. The halo of members is spread over a few parsecs around the star-forming sites. 

This differential distribution of mass accretion rates may be the product of a sequential star formation process: the youngest stars, likely the most actively accreting, are indeed expected to be found close to their birth sites, while dynamical evolution over a few Myr (the average age of the cluster) can explain displacements up to a few parsecs relative to the original location where stars were formed. If this is the case, the distribution of $\dot{M}_{acc}$ also reflects the presence of an intrinsic age spread within the region, hence corroborating the interpretation in terms of a time evolution over different ages. 

We attempted to pursue further this scenario by investigating the spatial distributions of distinct, equal-sized groups of objects at the two age extremities in Fig.\,\ref{fig:macc_age}. If our interpretation is correct, we would expect the presumed younger member group to be spatially more compact than the older one. A statistical method to probe spatial compactness is the minimum spanning tree (MST) method \citep{prim1957}. This (see also \citealp{allison2009, parker2011}) consists in computing the minimum path-length required to connect all points of a given sample in a network of direct point-to-point links with no closed loops. Shorter MST lengths will thus identify, to a certain degree of confidence, more compact populations. The significance of the result can be tested by randomly selecting two groups of objects of the same size and belonging to the same population as the two groups of objects under study, and computing the MST of the first random group, the MST of the second random group and the ratio between the two. The iteration of this procedure allows us to reconstruct the normal distribution of MST$_1$/MST$_2$ values when no significant difference exists between the spatial distributions of two given groups of objects selected among the population of interest. 

For our population of accreting objects, sampled in random groups of 20 to 30 elements, we derived a normal distribution of MST ratios centered on 1.0 (homogeneous spatial distribution) with a $\sigma$ of 0.2. Little evidence for an actual difference in the spatial distributions of ``younger'' vs. ``older'' members in Fig.\,\ref{fig:macc_age} can be inferred when comparing the result of their MST test to this distribution. Somewhat smaller values than the mean of 1.0 are obtained from the MST$_{younger}$/MST$_{older}$ ratio, oscillating around a significance of 1.5\,$\sigma$ that this difference is truly meaningful. This might suggest that the younger population is to some extent more compact than the older population; however, no conclusive evidence can be drawn from this analysis, and the small statistics at lower and higher ages in Fig.\,\ref{fig:macc_age} yields some dependence of the specific result on the specific selection of objects for the two groups. On the other hand, the complexity of the cluster, with its clumpy structure, the presence of at least two separated poles of star formation, the association with several massive OB stars, intrinsically translates to a non-linearity in the cluster evolution and, consequently, to a certain degree of ambiguity in the information provided by the spatial distribution of more evolved members.

\section{Conclusions} \label{sec:conclusions}
We have performed an extensive photometric investigation of accretion in the young open cluster NGC~2264. Our sample, homogeneous in the dataset and in the characterization of individual properties, includes $\sim$750 members, spanning in mass the range $\sim$0.1--2 M$_\odot$; of these, about 40\% show ongoing accretion. Accretion activity has been traced from the detection and monitoring of the $u$-band excess emission relative to non-accreting young stars; we showed that this excess is a reliable indicator of accretion, matching the properties observed at different wavelengths (H$_\alpha$ emission, IR excess), with the advantage of providing alone a direct proxy to the accretion luminosity. 

We re-examined the membership of the cluster from an accretion-related perspective, exploring a wider area around the spatial distribution of currently known members within the star-forming region. This allowed us to expand the census of accreting members, with the identification of 50 new candidates based on their accretion signatures at short wavelengths. We detected UV color excess ranging from a few tenths of a mag to $\sim$3 mag, with a sensitivity of $\sim$0.2 mag deriving from the definition of the reference color locus from the properties of non-accreting young stars. Following a new empirical prescription for the conversion of the UV excess into accretion luminosity, we inferred a complete picture of accretion in the cluster and monitored its variability over a few rotational cycles. A broad interval of $\dot{M}_{acc}$ values has been inferred within the region, ranging from $\sim$$10^{-7}$ to a few $\times10^{-10}$ M$_\odot$/yr; at any given stellar mass M$_*$, a large spread in $\dot{M}_{acc}$ values, typically extending over 2 orders of magnitude, is observed.

The average $\dot{M}_{acc}$ shows a statistically significant correlation with M$_*$; more uncertain is the functional form of this dependence, which in turn appears to be mass-dependent and likely time-evolving, as highlighted by the results inferred from previous studies for various young stellar populations. A statistical slope of 1.4$\pm$0.3 for the $\log{\dot{M}_{acc}}$ over $\log$ M$_*$ relationship is here derived for NGC~2264 in the mass sample probed; while the absence of the lower-mass ($<$0.1 M$_\odot$) stellar component may contribute to explain this less steep dependence than the value of $\sim$2 proposed in several studies, the large spread of $\dot{M}_{acc}$ values observed at a given mass attests the mere indicative value of such a relationship, which weighs the contributions of several trends as well as of undetected regimes.

Little contribution to the spread in the $\dot{M}_{acc}$ vs. M$_*$ distribution arises from $\dot{M}_{acc}$ variability, which is found to amount to $\sim$0.5~dex on the average, at least on days to months timescales; hence, this spread is physically relevant and provides valuable insight on the complexity of the accretion and evolutionary picture for a given stellar population. A differential distribution of evolutionary stages among members, such as we may deduce from the HR diagram of the cluster and/or from disk evolution tracers ($\alpha_{IRAC}$), would naturally impact the global accretion picture we may infer for the cluster as a whole. Indeed, $\dot{M}_{acc}$ is shown to anticorrelate with such intrinsic/evolutionary age indicators, although these trends are in turn not exempt from a large amount of scatter. The detected spread in $\dot{M}_{acc}$ may thus reflect the dynamics of cluster evolution and the presence of a non-negligible age spread among its members, be it the result of sequential star formation episodes and/or of a multiplicity of evolutionary patterns. Moreover, a diversity in the accretion regimes at play, from episodic, short-lived accretion bursts to steadier, funnel-flow accretion, may concur to shape the features of different stellar groups and result in a range of observed accretion rates, from the upper envelope of the $\dot{M}_{acc}$ vs. M$_*$ distribution to lower accretion rate regimes.

An important contribution toward the achievement of a complete physical picture of the accretion process is provided by a detailed analysis of the variability properties. Geometric effects constitute an important filter to the effective variability profile for a given system, contributing up to 75\%, on the average, of the detected $\dot{M}_{acc}$ variability; however, much information on the underlying physics is conveyed by the specific photometric signatures of this variability, and a comparative analysis over intrinsically different stellar groups may be of remarkable interest to discern different scenarios. This aspect will be fully addressed for the young stellar population of NGC~2264 in a forthcoming paper (Venuti et al., in prep.); this is intended to provide a complete characterization of the accretion process and its impact onto the dynamics of the cluster.

\begin{acknowledgements}
We would like to thank Christian Veillet, former director of CFHT, for granting discretionary time to perform the mapping survey in December 2010, and the Terapix center at Institut d'Astrophysique de Paris, and in particular Yannick Mellier, for the prompt processing of the MegaCam images obtained during this run. We also thank Nadine Manset and Jim Thomas at CFHT for efficient run scheduling and data retrieval procedures. We warmly thank Luisa Rebull for retrieving for us the L$_{acc}$ values she derived for her 2002 paper on disk-bearing objects in NGC~2264. We thank Kevin Covey for discussions on SDSS dwarf sequences and Lynne Hillenbrand for discussions on bolometric corrections scales; we also acknowledge useful discussion on the $\dot{M}_{acc}$--M$_*$ relationship with Beate Stelzer. This publication makes use of data products from the Sloan Digital Sky Survey and the Two Micron All Sky Survey. This project was in part supported by the grant ANR 2011 Blanc SIMI5-6 020 01. SHPA and APS acknowledge support from CNPq, CAPES and Fapemig.
\end{acknowledgements}

\bibliographystyle{aa}

\appendix

\section{Accretion-oriented revision of the census of NGC~2264 members: new CTTS candidate members and possible field contaminants}\label{app:tab_memb}

\begin{table*}
\caption{List of new candidate members from the accretion-based census of the population of NGC~2264.}
\label{tab:cand}
\centering
\begin{tabular}{l c c c c}
\hline \hline
  \multicolumn{1}{c}{Mon-ID} &
  \multicolumn{1}{c}{RA} &
  \multicolumn{1}{c}{Dec} &
  \multicolumn{1}{c}{Selection criteria} &
  \multicolumn{1}{c}{Additional information\tablefootmark{1}} \\
\hline
  Mon-001428 & 100.19257 & 9.71302 & UV, var\,I & {likely CTTS based on IRAC light curve}\\
  Mon-005009 & 100.54131 & 9.79834 & var\,II, lc & {irregular IRAC/CoRoT light curves}\\
  Mon-005278 & 100.64755 & 9.43207 & UV, var\,II & {IR excess}\\
  Mon-005326 & 100.62129 & 9.44550 & UV &\\
  Mon-005385 & 100.35425 & 9.75632 & UV, var\,I & {flat IRAC light curve}\\
  Mon-005455 & 100.21510 & 9.37313 & UV & {flat IRAC light curve}\\
  Mon-005596 & 100.51573 & 9.45916 & var\,I & {flat IRAC light curve}\\
  Mon-005664 & 100.22703 & 9.15885 & UV & {spotted-like CoRoT light curve}\\
  Mon-005745 & 100.51810 & 9.16134 & UV &\\
  Mon-005807 & 100.28199 & 9.23241 & UV, var\,I &\\
  Mon-005836 & 100.37121 & 9.30430 & UV & {flat IRAC light curve}\\
  Mon-006037\tablefootmark{*} & 99.87232 & 9.72772 & var\,II, lc & {spotted-like CoRoT light curve}\\
  Mon-006079 & 99.87132 & 9.71070 & UV, var\,I, var\,II & {eclipsing binary/very short period binary}\\
  Mon-006144 & 99.92838 & 9.55441 & UV & {flat IRAC light curve}\\
  Mon-006183 & 99.97203 & 9.82939 & UV, var\,II & {flat IRAC light curve}\\
  Mon-006286 & 100.12552 & 9.19233 & UV &\\
  Mon-006324 & 99.95310 & 9.29308 & UV &\\
  Mon-006325 & 100.06032 & 9.22705 & UV, var\,II &\\
  Mon-006369 & 99.87573 & 9.29094 & varI, var\,II &\\
  Mon-006398 & 99.81027 & 9.61444 & UV, var\,II & {flat IRAC light curve}\\
  Mon-006409 & 99.81342 & 9.64136 & UV & {flat IRAC light curve}\\
  Mon-006429 & 99.79541 & 9.53839 & UV, var\,I, var\,II & {strong IR excess; IRAC light curve of YSO}\\
  Mon-006465 & 99.85484 & 9.54391 & UV, var\,I, var\,II & {eclipsing binary}\\
  Mon-006491 & 99.85625 & 9.52760 & UV, var\,II, lc & {IR excess; CoRoT light curve of CTTS}\\
  Mon-006498 & 99.82751 & 9.39888 & UV & {flat IRAC light curve}\\
  Mon-006515 & 99.77643 & 9.66813 & UV &\\
  Mon-006873 & 99.79239 & 9.15089 & var\,I, var\,II &\\
  Mon-006902 & 99.75107 & 9.23571 & var\,I, var\,II &\\
  Mon-006930 & 99.76703 & 9.27055 & var\,II, lc & {likely CTTS based on CoRoT light curve}\\
  Mon-006950 & 99.77920 & 9.85195 & UV, var\,II &\\
  Mon-006985 & 99.77939 & 10.11909 & UV, var\,II &\\
  Mon-006986 & 99.79149 & 10.08689 & UV, var\,I, var\,II, lc & {likely CTTS based on CoRoT light curve}\\
  Mon-006991 & 99.84169 & 10.10648 & var\,II, lc & {likely CTTS based on CoRoT light curve; IR excess}\\
  Mon-006999 & 99.86647 & 10.06352 & var\,I &\\
  Mon-007004 & 99.85813 & 10.08544 & UV &\\
  Mon-007018 & 99.97253 & 10.10931 & var\,I, var\,II &\\
  Mon-007402 & 100.62534 & 9.94183 & var\,I &\\
  Mon-007418 & 100.57599 & 10.05188 & UV &\\
  Mon-007461 & 100.24085 & 10.11788 & UV, var\,II & {probable MIPS 24\,$\mu$m excess}\\
  Mon-007496 & 100.51515 & 10.08836 & var\,I, var\,II &\\
  Mon-007845 & 99.89187 & 9.40008 & UV &\\
  Mon-008139 & 99.75653 & 10.07972 & UV &\\
  Mon-008140 & 99.80434 & 9.61337 & UV, var\,II &\\
  Mon-008141 & 100.66490 & 9.67704 & UV &\\
  Mon-008142 & 99.76824 & 9.69662 & UV, var\,I, var\,II &\\
  Mon-014132 & 99.76480 & 9.27109 & var\,II, lc & {AA Tau-type CoRoT light curve; IR excess}\\
  Mon-020662 & 100.21076 & 10.03841 & UV &\\
  Mon-021771 & 100.51792 & 9.57561 & var\,I, var\,II & {flat IRAC light curve}\\
  Mon-025015 & 99.96704 & 9.40013 & UV &\\
  Mon-025101 & 99.85223 & 9.92528 & UV, var\,II &\\
  Mon-124054 & 100.39085 & 9.92262 & UV, var\,I &\\
\hline
\end{tabular}
\tablefoot{Selection criteria: UV = UV excess; var\,I = multi-year variability; var\,II = mid-term variability (i.e., on week timescales); lc = light curve morphology.\\
\tablefoottext{1}{{From Spitzer IRAC and/or CoRoT monitoring obtained during the CSI~2264 campaign (and/or additional IR photometry), when available.}} \\
\tablefoottext{*}{WTTS candidate member}
}
\end{table*}

\begin{table*}[h]
\caption{List of NGC~2264 members, with no apparent accretion activity from previous surveys, reclassified as CTTS from this accretion-based census of the population of NGC~2264.}
\label{tab:reclass}
\centering
\begin{tabular}{c c c c c}
\hline
\hline
  \multicolumn{1}{c}{Mon-ID} &
  \multicolumn{1}{c}{RA} &
  \multicolumn{1}{c}{Dec} &
  \multicolumn{1}{c}{Selection criteria} &
  \multicolumn{1}{c}{{Additional information\tablefootmark{1}}} \\
\hline
  Mon-000293 & 100.25166 & 9.68758 & UV & \\
  Mon-000316 & 100.23172 & 9.64025 & UV & {IR excess; likely CTTS based on IRAC light curve}\\
  Mon-000340 & 100.48304 & 9.67968 & UV & {flat IRAC light curve}\\
  Mon-000507 & 100.26089 & 9.85702 & UV & {flat IRAC light curve; possible 8 $\mu$m excess}\\
  Mon-000685 & 100.18949 & 9.81950 & UV & {flat IRAC light curve}\\
  Mon-000718 & 100.26733 & 9.34567 & UV & {flat IRAC light curve}\\
  Mon-000750 & 100.32659 & 9.52487 & UV & {flat IRAC light curve; possible 8 $\mu$m excess}\\
  Mon-000823 & 100.20891 & 9.95111 & UV\\
  Mon-000914 & 100.21629 & 9.63220 & UV & {very strong IR excess}\\
  Mon-000957 & 100.19014 & 9.20354 & UV & {IR excess}\\
  Mon-000959 & 100.18845 & 9.26850 & var & {IR excess}\\
  Mon-001060 & 100.12844 & 10.00076 & UV & {IR excess; likely CTTS based on IRAC light curve}\\
  Mon-001211 & 99.95979 & 9.52143 & UV & {IR excess}\\
  Mon-001224 & 100.06953 & 9.67295 & UV\\
  Mon-001241 & 100.14078 & 9.81216 & UV & {IR excess; likely CTTS based on IRAC light curve}\\
  Mon-001267 & 100.14299 & 9.42142 & var & {strong IR excess; IRAC light curve of YSO}\\
  Mon-001280 & 99.91768 & 9.30604 & UV & {strong IR excess}\\
  Mon-001576 & 100.13291 & 10.11898 & UV & {strong MIPS 24 $\mu$m excess}\\
  Mon-001582 & 100.49939 & 10.06379 & UV & {strong IR excess}\\
\hline\end{tabular}
\tablefoot{Selection criteria: UV = UV excess; var = multi-year variability. \\
\tablefoottext{1}{{From Spitzer/IRAC monitoring obtained during the CSI~2264 campaign (and/or additional IR photometry), when available.}} 
}
\end{table*}

\begin{table*}
\caption{Objects with questioned membership from the CFHT UV-census.}
\label{tab:questioned}
\centering
\begin{tabular}{l c c c c c c c c}
\hline\hline
  \multicolumn{1}{c}{Mon-ID} &
  \multicolumn{1}{c}{RA} &
  \multicolumn{1}{c}{Dec} &
  \multicolumn{1}{c}{$u$} &
  \multicolumn{1}{c}{$g$} &
  \multicolumn{1}{c}{$r$} &
  \multicolumn{1}{c}{$i$} &
  \multicolumn{1}{c}{Membership criteria} \\
\hline
  Mon-000003 & 100.43878 & 9.78583 & 20.460 & 18.204 & 16.924 & 16.206 & P, X\\
  Mon-000004 & 100.56094 & 9.84104 & 17.808 & 15.998 & 15.075 & 14.594 & P, RV\\
  Mon-000025 & 100.51118 & 9.97441 & 18.025 & 16.134 & 15.193 & 14.690 & P, RV\\
  Mon-000031 & 100.55471 & 9.76763 & 17.801 & 16.118 & 15.307 & 14.888 & P, RV\\
  Mon-000387 & 100.45535 & 9.68496 & 20.530 & 18.408 & 17.273 & 16.681 & X\\
  Mon-000414 & 100.55863 & 9.59579 & 18.446 & 16.440 & 15.431 & 14.944 & P, RV\\
  Mon-000593\tablefootmark{*} & 100.25704 & 9.35157 & 17.793 & 15.822 & 15.077 & 14.836 & P, RV\\
  Mon-000600\tablefootmark{*} & 100.20549 & 9.98095 & 22.567 & 19.684 & 18.106 & 17.212 & X\\
  Mon-000696\tablefootmark{*} & 100.33708 & 9.57170 & 20.309 & 19.512 & 19.589 & 18.516 & h\\
  Mon-000820 & 100.48089 & 9.50528 & 19.538 & 17.513 & 16.490 & 15.902 & X\\
  Mon-000944 & 100.45795 & 9.68479 & 20.258 & 18.274 & 17.185 & 16.629 & X\\
  Mon-000946\tablefootmark{*} & 100.20165 & 9.99424 & 21.252 & 18.810 & 17.472 & 16.715 & II\\
  Mon-000966 & 100.19639 & 9.30923 & 17.729 & 15.882 & 14.942 & 14.415 & P, RV\\
  Mon-000976 & 100.43282 & 9.15588 & 21.325 & 19.108 & 17.798 & 17.058 & X\\
  Mon-000983 & 100.24170 & 9.30085 & 16.047 & 14.663 & 14.206 & 14.042 & RV\\
  Mon-000988 & 100.08816 & 9.98401 & 20.823 & 18.249 & 16.758 & 15.930 & P, RV\\
  Mon-001043\tablefootmark{*} & 100.06439 & 9.71678 & 21.746 & 19.149 & 17.695 & 16.854 & P, II/III\\
  Mon-001046 & 99.98493 & 9.72542 & 15.811 & 14.455 & 14.048 & 13.881 & RV\\
  Mon-001086 & 99.88986 & 9.86127 & 18.279 & 16.524 & 15.639 & 15.144 & P, RV\\
  Mon-001192 & 100.16157 & 9.36066 & 17.954 & 15.866 & 15.079 & 14.799 & RV\\
  Mon-001207 & 99.98039 & 9.98644 & 21.113 & 18.345 & 16.911 & 16.009 & h\\
  Mon-001283\tablefootmark{*} & 99.86101 & 9.51055 & 21.364 & 18.750 & 17.590 & 17.012 & RV\\
  Mon-001289 & 99.85268 & 9.37588 & 16.778 & 15.082 & 14.379 & 14.084 & P, RV\\
  Mon-001293 & 99.89724 & 9.54233 & 18.331 & 16.315 & 15.239 & 14.681 & P, RV\\
  Mon-001311 & 99.81513 & 9.65357 & 18.014 & 16.134 & 15.181 & 14.684 & P, RV\\
  Mon-001393 & 100.55236 & 9.98531 & 17.072 & 15.373 & 14.609 & 14.195 & P, RV\\
  Mon-006019\tablefootmark{*} & 100.14231 & 9.42022 & 19.511 & 17.507 & 16.217 & 15.507 & P\\
  Mon-007454 & 100.12886 & 10.08013 & 17.799 & 16.099 & 15.494 & 15.140 & P\\
\hline\end{tabular}
\tablefoot{Membership criteria, in the literature, for the listed objects: P = photometric member (i.e., located on a cluster sequence in the R vs. R-I diagram); X = X-ray emitter, from \citet{sung09}; RV = radial velocity member, from \citet{furesz06}; h = H$_\alpha$ emission candidate, from \citet{sung09}; II = Class II object, from \citet{sung09}; II/III = Class II/III object, from \citet{sung09}.\\
\tablefoottext{*}{More uncertain cases}\\
}
\end{table*}

\end{document}